\documentclass[10pt,journal,compsoc]{IEEEtran}

\newcommand{\ignore}[1]{}
\newcommand{\sagpp}{{\sc SparCE}}
\usepackage[pass]{geometry}
\usepackage{fancyhdr}
\usepackage[normalem]{ulem}
\usepackage[hyphens]{url}
\usepackage{hyperref}

\usepackage[]{graphicx}
\usepackage{stfloats}  
\usepackage{cite}  
\usepackage{psfrag}  
\usepackage{subfigure}
\usepackage{wrapfig}
\usepackage{epsfig}
\usepackage{url}
\usepackage{amsmath, amssymb}
\usepackage{array}
\usepackage{times}
\usepackage{multicol}
\usepackage{algorithm}
\usepackage{algorithmic}
\usepackage{mathrsfs}
\usepackage{multirow}
\usepackage[normalem]{ulem}

\usepackage[utf8x]{inputenc} 
\usepackage{ucs} 
\usepackage{amsfonts} 
\usepackage{makeidx} 
\usepackage[dvipsnames,usenames]{color} 
\usepackage{array} 
\usepackage{colortbl} 
\usepackage{booktabs} 
\usepackage[inline]{enumitem}

\begin{document}
\title{\vspace{-0.3in}{\sc SparCE}: \underline{Spar}sity aware General Purpose \underline{C}ore \underline{E}xtensions to Accelerate Deep Neural Networks}
\author{\IEEEauthorblockN{Sanchari Sen, Shubham Jain, Swagath Venkataramani, Anand Raghunathan} \\
\IEEEcompsocitemizethanks{\IEEEcompsocthanksitem Sanchari Sen, Shubham Jain and Anand Raghunathan are with School of Electrical and Computer Engineering, Purdue University, West Lafayette, IN, 47906, USA. \protect\\
E-mail:  \{sen9,jain130,raghunathan\}@purdue.edu
\IEEEcompsocthanksitem Swagath Venkataramani is with IBM T.J Watson Research Center, Yorktown Heights, NY. \protect\\
E-mail:  swagath.venkataramani@ibm.com }
}

\IEEEtitleabstractindextext{
\begin{abstract}
\noindent
Deep Neural Networks (DNNs) have emerged as the method of choice for solving a wide range of machine learning tasks. Satiating the enormous growth in computational demand posed by DNNs is a key challenge for computing system designers and has most commonly been addressed through the design of custom accelerators. However, these specialized accelerators that utilize large quantities of multiply-accumulate units and on-chip memory are prohibitive in many design scenarios ({\em e.g.}, wearable devices and IoT sensors), due to the stringent area/cost constraints. Therefore, accelerating DNNs on these low-power systems, comprising of mainly the indispensable general-purpose processor (GPP) cores, requires new approaches. In this work, we focus on improving the performance of DNNs on GPPs by exploiting a key attribute of DNNs, \emph{i.e.} sparsity. We propose Sparsity aware Core Extensions (\sagpp) - a set of micro-architectural and ISA extensions that leverage sparsity and are minimally intrusive and low-overhead. We address the key challenges associated with exploiting sparsity in GPPs, \emph{viz.}, dynamically detecting whether an operand ({\em e.g.}, the result of a load instruction) is zero and subsequently skipping a set of future instructions that use it. To maximize performance benefits, our design ensures that the instructions to be skipped are prevented from even being \emph{fetched}, as squashing instructions comes with a penalty ({\em e.g.}, a pipeline stall). \sagpp~consists of 2 key micro-architectural enhancements. First, a Sparsity Register File (SpRF) is utilized to track registers that are zero. Next, a Sparsity aware Skip Address (SASA) table is used to indicate instruction sequences that can be skipped, and to associate specific SpRF registers to trigger instruction skipping. When an instruction is fetched, \sagpp~dynamically \emph{pre-identifies} whether the following instruction(s) can be skipped, and if so appropriately modifies the program counter, thereby skipping the redundant instructions and improving performance. We model \sagpp~using the gem5 architectural simulator, and evaluate our approach on 6 state-of-the-art image-recognition DNNs in the context of both training and inference using the Caffe deep learning framework. On a scalar microprocessor, \sagpp~achieves 1.11$\times$-1.96$\times$ speedups across both convolution and fully-connected layers with 10\%-90\% sparsity, and 19\%-31\% reduction in execution time at the overall application-level. We also evaluate \sagpp~on a 4-way SIMD ARMv8 processor using the OpenBLAS library, and demonstrate that \sagpp~achieves 8\%-15\% reduction in the application-level execution time.
\end{abstract}
}

\maketitle
\IEEEdisplaynontitleabstractindextext

\section{Introduction}
\label{sec:introduction}
\noindent
Deep neural networks (DNNs) have revitalized the field of machine learning by achieving accuracy levels beyond human perception in a variety of image, video, text and speech processing tasks~\cite{resNet,vgg16,googleNet,Venugopalan:2015,Hannun:2014,Zhang:2015,Hinton:2012,Li:2016}. Today, they have emerged as the \emph{de facto} solution approach for challenging artificial intelligence problems and are deployed in several products and services, including Google's image and voice search~\cite{Google:2013}, Facebook's DeepFace~\cite{deepface} and DeepText~\cite{Facebook:2016}, Apple's Siri~\cite{Apple:2016}, to name a few. However, one of the key challenges to the ubiquitous adoption of DNNs is their computational demand, which far outstrips the capabilities of modern computing platforms. For example, ResNet-152~\cite{resNet}, a state-of-the-art DNN for image recognition, requires $\sim$11.3 giga operations to classify a single 224$\times$224 image. As DNNs begin to pervade the spectrum of computing devices from high performance servers to low-power edge/IoT devices, the demands on compute efficiency are expected grow even more stringent. Consequently, there is an urgent need to explore new approaches to improve their implementation efficiency.

Realizing this need, prior research efforts have explored several key directions. A majority of research efforts exploit the compute and communication characteristics of DNNs to efficiently parallelize them on distributed platforms~\cite{Krizhevsky:2014,Iandola:2015,Dean:2012,Rhu:2016,Das:2016,Zlateski:2016,Nervana:2016} or design specialized accelerator architectures~\cite{Farabet:2011,Chakradhar:2010,Chen:2014,Chen:2014:1,Venkataramani:2017,Jouppi:2017,Reagen:2016,Venkataramani:2013,Eldridge:2015,Majumdar:2012,Gokhale:2014,eyeriss}. Other efforts leverage the intrinsic error resilience of DNNs to approximate selected computations~\cite{Gupta:2015,Venkataramani:2014,Zhu:2016,Courbariaux:2015,deepComp,Jaderberg:2014,Seide:2014,Zhang:2015:1} or utilize non-CMOS technologies whose device characteristics match the compute kernels of DNNs~\cite{Liu:2015,Ramasubramanian:2014,Shafiee:2016}.

Complementary to the above efforts, we focus on a different design scenario and target deeply embedded systems such as wearables and IoT edge devices where the additional area/ cost imposed by custom-accelerators is prohibitive. For example, even a low power DNN accelerator such as Eyeriss~\cite{eyeriss} occupies $12.25mm^2$ area which is 30$\times$ larger than the $0.4mm^2$ occupied by an ultra low power Cortex A35 core~\cite{arm}. Accelerating DNNs on these low-power systems then comes down to accelerating them on the indispensable general-purpose processor (GPP) cores already present in these systems. We focus on improving the execution time of DNNs on GPPs by leveraging \emph{sparsity} in different DNN data-structures, \emph{viz.}, features, weights and backpropagated errors. Sparsity in DNNs can be both static or dynamic depending on whether the zero values remain constant or vary across different inputs to the network. Sparsity in weights, introduced by pruning connections in the network after training, is static in nature. In contrast, feature and error sparsities, caused by the thresholding nature of the ReLU (Rectified Linear Unit) activation functions, are dynamic in nature. 

\begin{figure}[htb]
	\centering
	\includegraphics[clip, width=1.0\columnwidth]{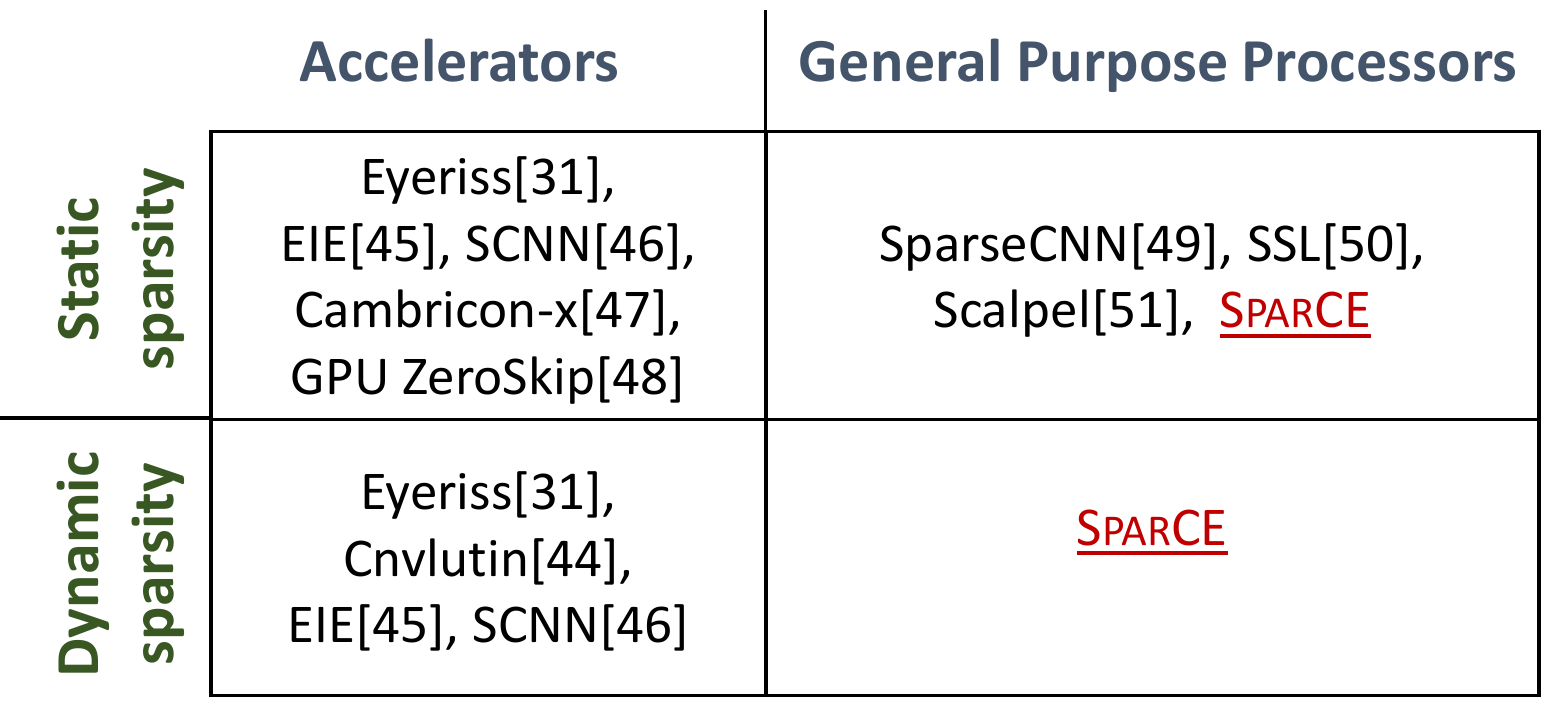}
	\caption{Related work: Exploiting sparsity in DNNs}
	\label{fig:relwork}
\end{figure}

Prior efforts that exploit sparsity to accelerate DNNs can be grouped into two classes based on whether they exploit static or dynamic sparsity, as shown in Figure~\ref{fig:relwork}. Specialized accelerators~\cite{cnvlutin, eyeriss, eie, scnn, cambriconx, Park:2016} have been proposed to exploit both forms of sparsity. However, these techniques are closely tied to a given accelerator architecture thereby preventing their applicability to GPPs. In contrast, static sparsity in weights can be exploited on GPPs through software-only approaches~\cite{BLiu:2015,Wen:2016,Yu:2017} that involve sparse encodings and sparse matrix multiplication routines. However, extending them to exploit the intermediate levels of dynamic sparsity (40\%-70\%) can be counter-productive because of the encoding overhead involved at runtime. For example, the sparse encoding overhead involved in performing sparseBLAS based matrix multiplication on the sparse conv3 layer features of AlexNet~\cite{alexNet} causes a slowdown of 2.78$\times$. 
We observe across 6 image-recognition DNNs that dynamic sparsity in features results in 45.1\% of the computations being rendered redundant during inference, presenting a significant opportunity for improving performance. We believe our effort, \sagpp, is the first to successfully exploit dynamic sparsity in DNNs on GPPs. To that end, we propose micro-architectural and ISA extensions that are minimally intrusive and low-overhead. Moreover, since static sparsity is a special case of dynamic sparsity where the location of zeros remains constant across inputs, the extensions also allow \sagpp~to exploit static sparsity in weights. 

To exploit dynamic sparsity, GPPs need to be equipped to dynamically detect if the result of an instruction ({\em e.g.}, a load from a sparse data structure) is zero and if so, skip a set of future instructions that use its result. However, there are two key challenges: (i) the instructions to be skipped often do not immediately follow the instruction that determines whether they may be; moreover, the instructions to be skipped may be non-contiguous in the program, and (ii) to maximize performance benefits, the instructions to be skipped should be \emph{prevented from even being fetched}, as squashing the instruction in-flight would diminish the benefits by introducing a pipeline stall.

 \vspace{5pt}
{\bf \noindent Sparsity aware Core Extensions (\sagpp).} To overcome the aforementioned challenges, we propose \underline{Spar}sity aware \underline{C}ore \underline{E}xtensions (\sagpp), comprised of two key micro-architectural enhancements. First, \sagpp~contains a \emph{Sparsity Register File} (SpRF) to track which general purpose registers that contain a zero. We achieve this by augmenting the writeback stage of the processor to check if the update to a register is zero and appropriately modify the corresponding SpRF entry. Next, a \emph{Sparsity Aware Skip Address} (SASA) table is used to store instruction sequences and the conditions under which they can be skipped \emph{i.e.,} the registers in the SpRF that need to be zero for the instructions to become redundant. Whenever \sagpp~fetches an instruction, it uses the SASA table and the SpRF to \emph{pre-identify} whether the following instruction(s) can be skipped. If so, the program counter is modified to directly fetch the next irredundant instruction. 

In summary, the key contributions of this work are:
\begin{itemize}
\item We propose Sparsity aware Core Extensions (\sagpp) to accelerate DNNs on general purpose processors by skipping redundant computations borne out of sparsity in the different DNN data-structures. 
\item \sagpp~comprises of micro-architectural enhancements to dynamically track when the result of an instruction is zero, pre-identify future instructions that are rendered redundant, and prevent them from being fetched and executed, thereby improving performance. 
\item We evaluate \sagpp~on a suite of 6 state-of-the-art image-recognition DNN benchmarks using the Caffe deep learning framework. We achieve application-level speedups of 19\%-31\% 
 on low-power embedded scalar microprocessors. We also achieve speedups of 8\%-15\% 
 over highly optimized baseline implementations that use OpenBLAS on an ARMv8 processor with 4-way SIMD and prefetching support.
\end{itemize}

The rest of the paper is organized as follows. Section~\ref{sec:motivation} explains the sources of sparsity in DNNs. Section~\ref{sec:design} details the key design principles of \sagpp~and demonstrates them in the context of an in-order pipelined processor. Section~\ref{sec:codeGen} shows \sagpp~in action using the ARM-BLAS GEMM routine as a case study. Section~\ref{sec:exptMeth} describes the experimental methodology and the results are presented in Section~\ref{sec:results}. Section~\ref{sec:relatedwork} presents related research efforts and Section~\ref{sec:conclusion} concludes the paper.

\vspace*{-0pt}
\section{Sparsity in DNNs: Sources and Opportunity}
\label{sec:motivation}
\noindent
In this section, we first provide a brief background on DNNs. We then explain the various static and dynamic sources of sparsity in the different DNN data-structures, and quantify the opportunity for performance improvement afforded by sparsity.

\begin{figure*}[htb]
  \centering
  \includegraphics[clip,width=\textwidth]{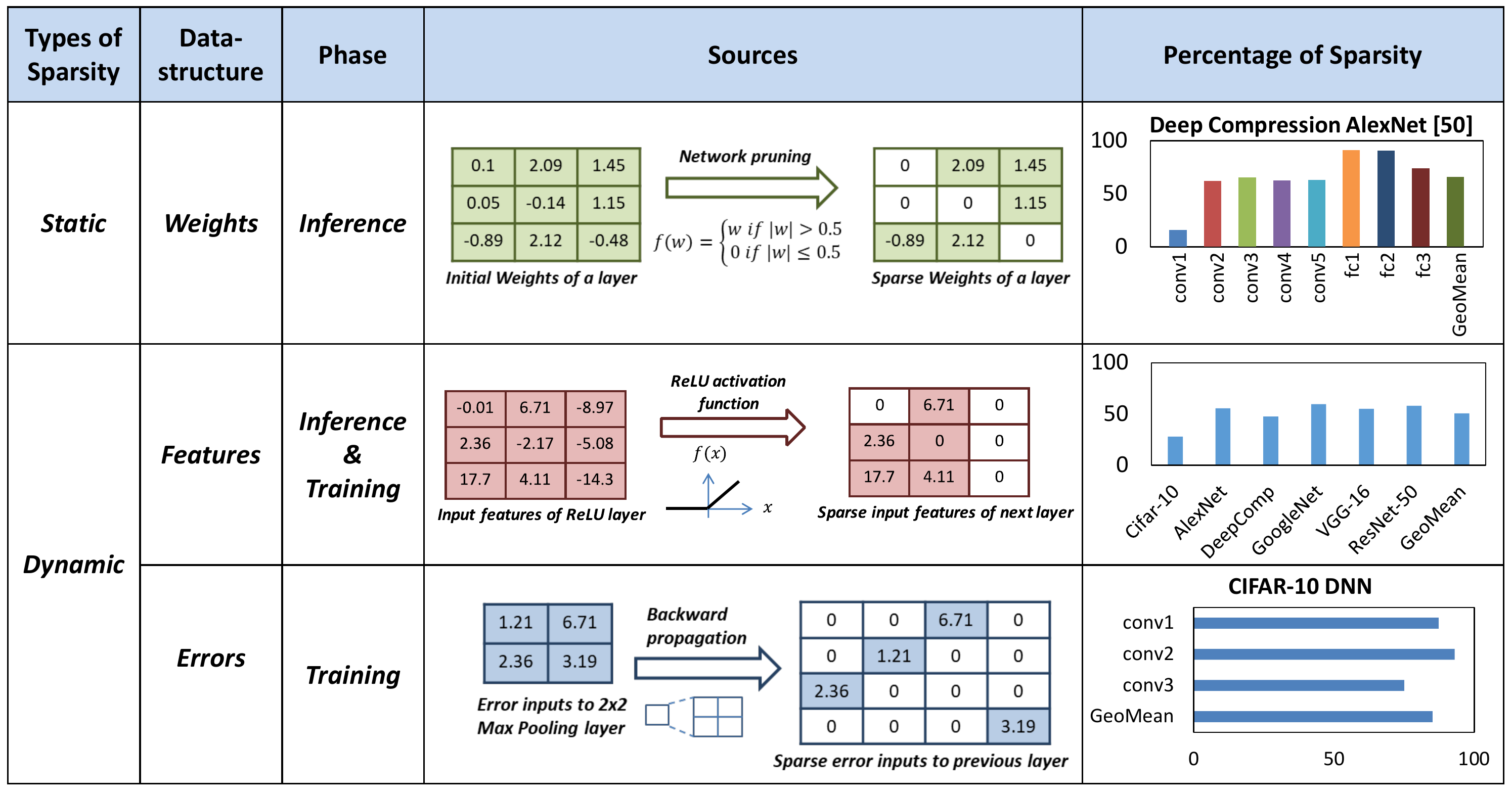}
  \vspace*{-20pt}
  \caption{Different forms of sparsity in DNNs}
  \label{fig:src_sparsity}
\end{figure*}
\subsection{DNN: Background}
DNNs are networks of primitive compute units called \emph{neurons}, organized into layers. Each layer is associated with a set of parameters called weights. DNNs operate in two phases \emph{viz.} training and inference. During the training phase, a labelled training dataset is used to iteratively refine the weights of the DNN. In the inference phase, the trained DNN is used to classify new inputs. 

Computationally, DNN executions iteratively perform 3 key steps \emph{viz.} Forward Propagation (FP), Backpropagation (BP), and Weight Gradient and Update (WG), that operate on 4 data-structures, \emph{viz.}, features, weights, errors and gradients. All three steps (FP, BP, WG) are performed during the training phase, while inference involves only the FP step. In FP, inputs to the DNN are propagated through its layers to produce the DNN outputs. In each layer, the input features are operated on with its weights to produce its output features, which are then fed to the next layer and so on. In BP, errors observed at the output of the DNN are propagated backwards through each layer of the DNN. In this case, the error at the output of a layer is operated on with its weights to compute the error at its inputs. In WG, the input features and the output error of each layer are used to refine its weights. 

\subsection{Sources of Sparsity in DNNs}
In practice, all major DNN data-structures - features, weights, errors and gradients - exhibit significant levels of sparsity, which can be exploited for computational savings. Three of the four data-structures (all except weight gradients) are used as inputs to multiply-and-accumulate operations in the different steps (FP/BP/WG), which become redundant when one of the input operands is zero. Among these three sparse multiply-and-accumulate operands, weights exhibit static sparsity that remains constant across different inputs while the remaining two data-structures (features and errors) exhibit dynamic sparsity that varies dynamically across different inputs. Figure~\ref{fig:src_sparsity} summarizes the sparsity in the different DNN data-structures, which we describe in the remainder of this subsection.

\subsubsection{Static Sparsity}

\textbf{Weight Sparsity.} Sparsity in weights occurs during the inference phase of the DNN. As shown in Figure~\ref{fig:src_sparsity}, after training, connections whose weights are close to zero are pruned to compress the model size~\cite{deepComp,Reagen:2016,Jaderberg:2014}. The last column in Figure~\ref{fig:src_sparsity} shows the fraction of zero weights in the different layers of the AlexNet model trained using deep compression~\cite{deepComp}. We find the sparsity to vary between 18\%-85\% across the different layers. Weight sparsity is static  in nature because of the fact that zero weights are identified before the inference phase.

\subsubsection{Dynamic Sparsity}

\begin{figure}[htb]
  \centering
  \vspace*{-5pt}
  \includegraphics[clip,width=0.9\columnwidth]{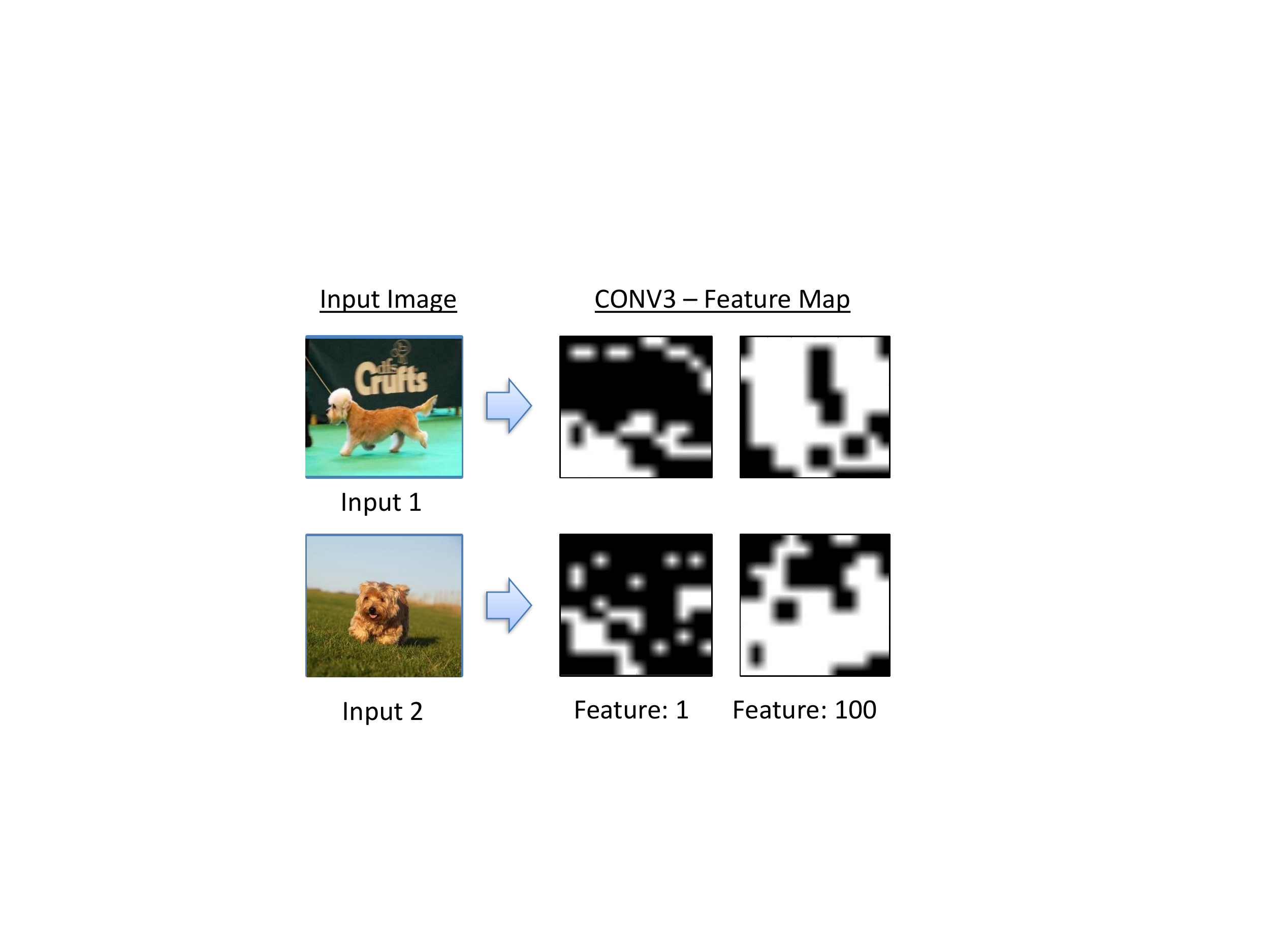}
  \caption{Variation in feature sparsity of AlexNet CONV3 layer across input images}
  \label{fig:ft_sparsity2}
\end{figure}

\textbf{Feature sparsity.} Sparsity in features stems from the thresholding nature of the activation function present at the output of each layer. As shown in Figure~\ref{fig:src_sparsity}, the predominantly used ReLU (Rectified Linear Unit) activation function clips negative inputs to zero. Figure~\ref{fig:src_sparsity} also shows the average feature sparsity exhibited by various DNN benchmarks, which ranges between $\sim$25\%-60\%. Feature sparsity has a \emph{dynamic} nature \emph{i.e.,} because neurons whose outputs are zero vary considerably across inputs. Figure~\ref{fig:ft_sparsity2} illustrates this property using feature maps obtained at the conv3 of AlexNet, produced by two different inputs from the ImageNet dataset. We find the spatial variation in white pixels, which indicate zero output, to be substantial across inputs.

\textbf{Error Sparsity.} Sparsity in the error data-structure originates from two sources. First, the derivative of the activation function, such as ReLU, is zero when the error at the output of the layer is negative. Next, when errors are propagated back through a max-pooling layer, as shown in Equation~\ref{eq:maxpool}, only one input of each pooling window is set a non-zero error value. 
\begin{equation}
\label{eq:maxpool}
\frac{\partial E}{\partial y_{l}} (x+p, y+q)= \begin{cases}
0, & \text{if } y_{l+1}(x,y)\neq y_{l}(x+p,y+q)\\
\frac{\partial E}{\partial y_{l+1}}, & \text{otherwise}
\end{cases}
\end{equation}

For example, if a pooling window of size 2$\times$2 is used, at least three quarters of the error values are sparse.
Similar to feature sparsity, the error sparsity is also dynamic. The average error sparsity in different layers of a CIFAR-10 DNN is shown in Figure~\ref{fig:src_sparsity}.

\subsection{Opportunity for Computational Savings}
\begin{wrapfigure}{r}{0.55\columnwidth}
  \centering
  \vspace*{-5pt}
  \includegraphics[clip,width=0.55\columnwidth]{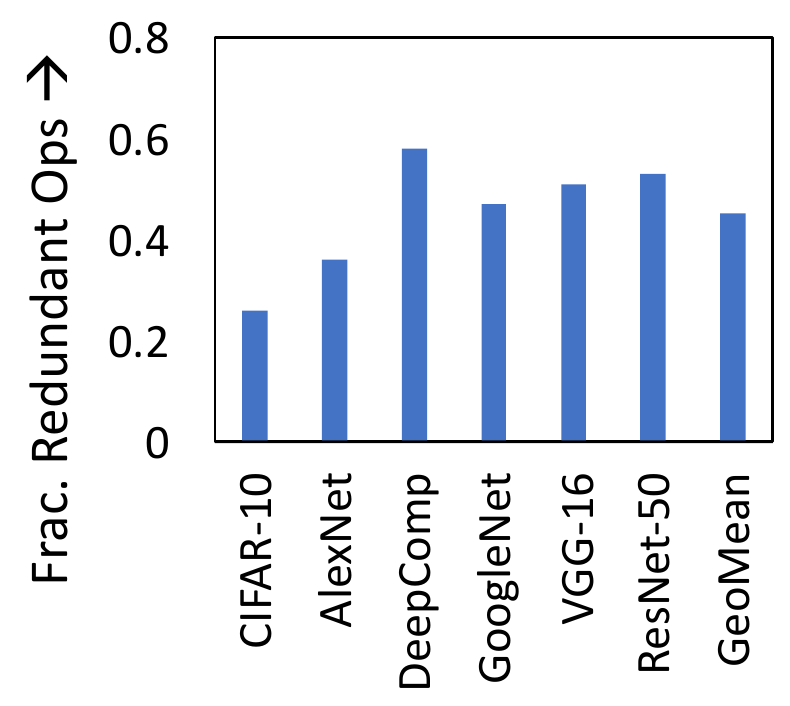}
  \vspace*{-17pt}
  \caption{Redundant ops} 
  \label{fig:bm_opportunity}
  \vspace*{-10pt}
\end{wrapfigure}
\begin{figure}[htb]
  \centering
  \vspace*{-10pt}
  \includegraphics[clip,width=0.9\columnwidth]{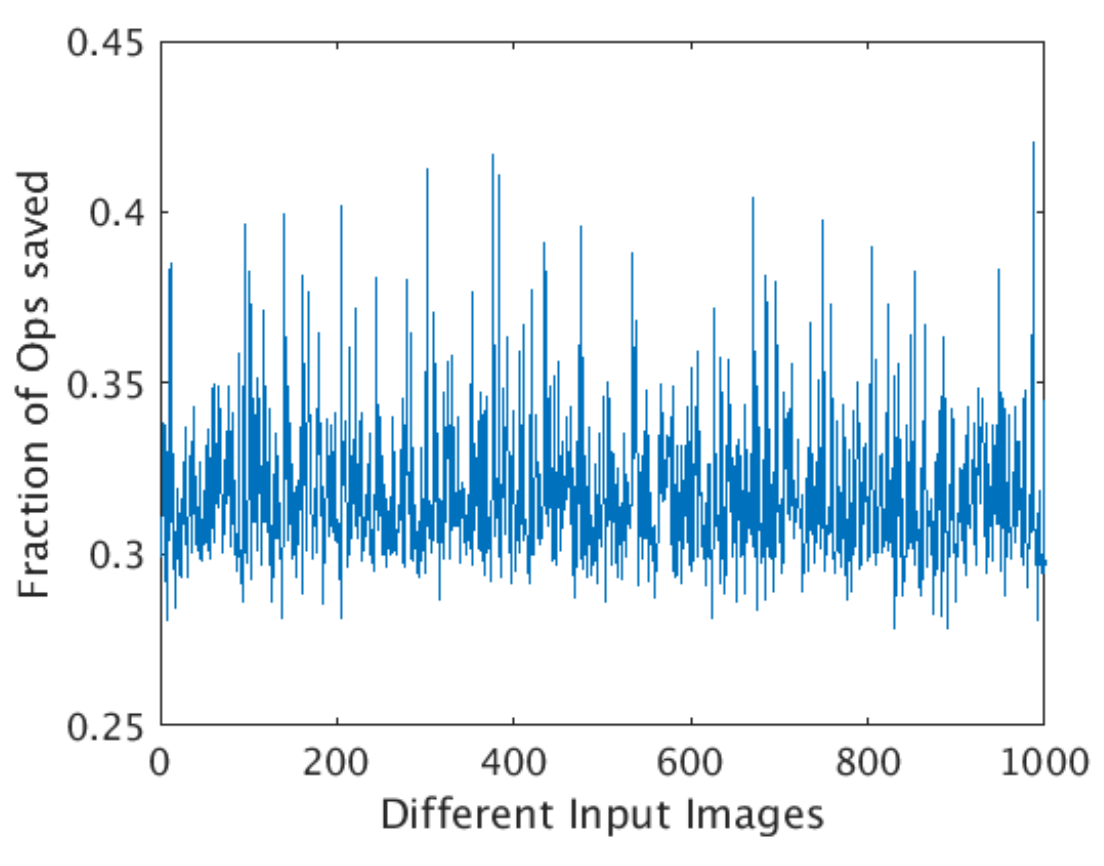}
  \caption{Fraction of redundant ops across different inputs of AlexNet}
  \label{fig:inp_var_sparsity}
\end{figure}

Figure~\ref{fig:bm_opportunity} shows the fraction of multiply-accumulate (MAC) computations that are rendered redundant for each DNN benchmark due to dynamic sparsity in features during inference. We find that between 25\%-60\% (average: 45\%) of the computations can be skipped, underscoring the substantial opportunity for performance improvement. Figure~\ref{fig:inp_var_sparsity} shows how the fraction of redundant operations varies across 1000 different inputs for the AlexNet DNN. We observe $\sim$14\% variation across inputs, although each input shows considerable opportunity for reduction in execution time (minimum: 28\%).

In summary, the dynamic sparsity present in the feature and error data-structures offers a substantial opportunity to accelerate DNNs. However, the levels of sparsity are not extreme enough to completely exploit them in software, and this coupled with their dynamic nature necessitates hardware solutions to realize benefits in the context of general purpose processors.

\section{SparCE: Sparsity Aware General \\ Purpose Core Extensions}
\label{sec:design}
\noindent
To exploit the different forms of sparsity and improve DNN performance on GPPs, we propose 
Sparsity aware Core Extensions (\sagpp), a set of micro-architectural and ISA extensions that are general-purpose, minimally intrusive and low-overhead. In this section, we present the key ideas behind \sagpp~and describe how they can be integrated within an in-order processor pipeline. 

\subsection{Challenges}
\begin{figure}[htb]
  \centering
  \vspace*{-5pt}
  \includegraphics[clip,width=1.0\columnwidth]{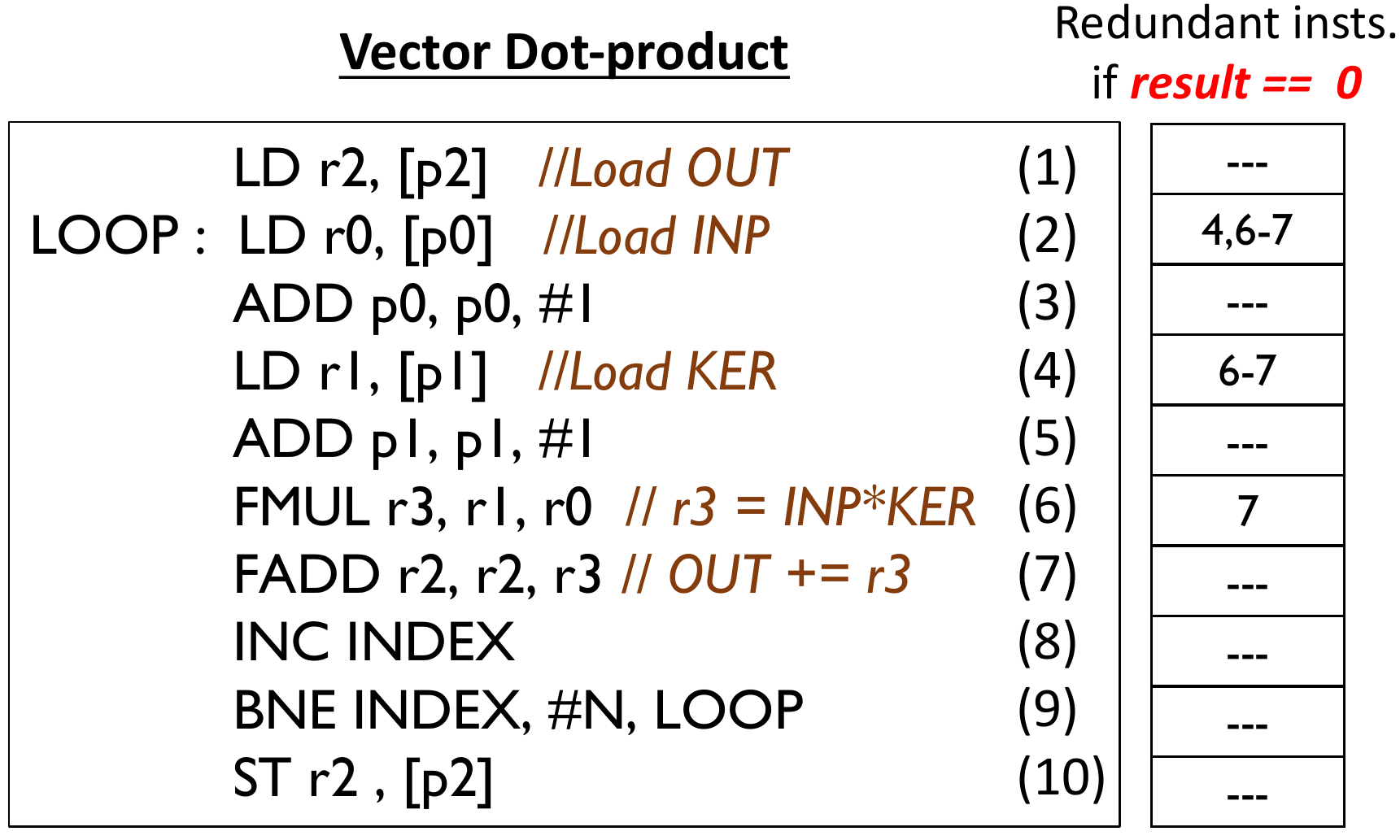}
  \caption{Redundant instructions due to sparsity in vector dot-product evaluation}
  \label{fig:loop}
\end{figure}

The key challenge in exploiting sparsity is to equip the processor with the ability to dynamically detect if the result of an instruction is zero and if so, skip a list of future instructions that are rendered redundant. We illustrate this challenge using the assembly code snippet shown in Figure~\ref{fig:loop}, which computes the dot-product of two vectors, $INP$ and $KER$, each of size $N$, to produce a scalar $OUT$. Registers $r0$, $r1$ and $r2$ hold the data operands, while $p0$, $p1$ and $p2$ are pointers that hold their respective memory locations. For each instruction in the program, Figure~\ref{fig:loop} shows the instructions that can be skipped when its result is zero. For example, when the $INP$ load returns a zero (Inst. 2), the subsequent $KER$ load (Inst. 4), and the multiply and add instructions can be skipped (Insts. 6-7). It is noteworthy that the computational savings is a weighted sum of the number of instructions skipped and the cycles taken by each instruction. For instance, floating point multiply and add instructions may take 3-5 cycles to execute, while a load incurs variable cycles depending on the level of cache hierarchy accessed.

The following observations highlight the challenges in detecting and benefiting from sparsity.
\begin{itemize}
\item \textbf{Location of redundant instructions.} In a program, the instructions that can be skipped may not immediately follow the instruction producing the zero result. Worse, redundant instruction sequences may be scattered non-contiguously through the program. For instance, in the program shown in Figure~\ref{fig:loop}, when inst. 2 returns a zero, 2 non-contiguous instruction sequences (Inst. 4 and Insts. 6-7) need to skipped. Hence, efficient ways to \emph{capture which instructions can be skipped} is key to leveraging sparsity. 
\item \textbf{Avoiding redundant instruction fetches.} To maximize performance, the \emph{instructions that can be skipped need to be prevented from even being fetched}. This is especially critical in the context of in-order pipelines, where even if the instruction is squashed after being fetched, it would introduce a bubble in the pipeline. For example, if the condition for $r0$ or $r1$ being zero is checked after the $FMUL$ instruction is fetched (Inst. 6), it would result in a bubble flowing through the pipeline in place of Inst. 6. It is worth noting that, in the context of multi-cycle instructions, squashing the instruction after it is fetched can still improve performance.
\item \textbf{Use of SIMD instructions.} GPPs use vector units with SIMD (Single Instruction, Multiple Data) execution engines to exploit fine-grained data parallelism in workloads. SIMD instructions can be skipped only if computations performed on all the vector lanes are redundant. This constrains the sparsity to be relatively coarse grained, as irregularly scattered zero values are not beneficial.
\end{itemize}

\subsection{SparCE: Overview}
\begin{figure}[htb]
  \centering
  \vspace{-5pt}
  \includegraphics[clip,width=\columnwidth]{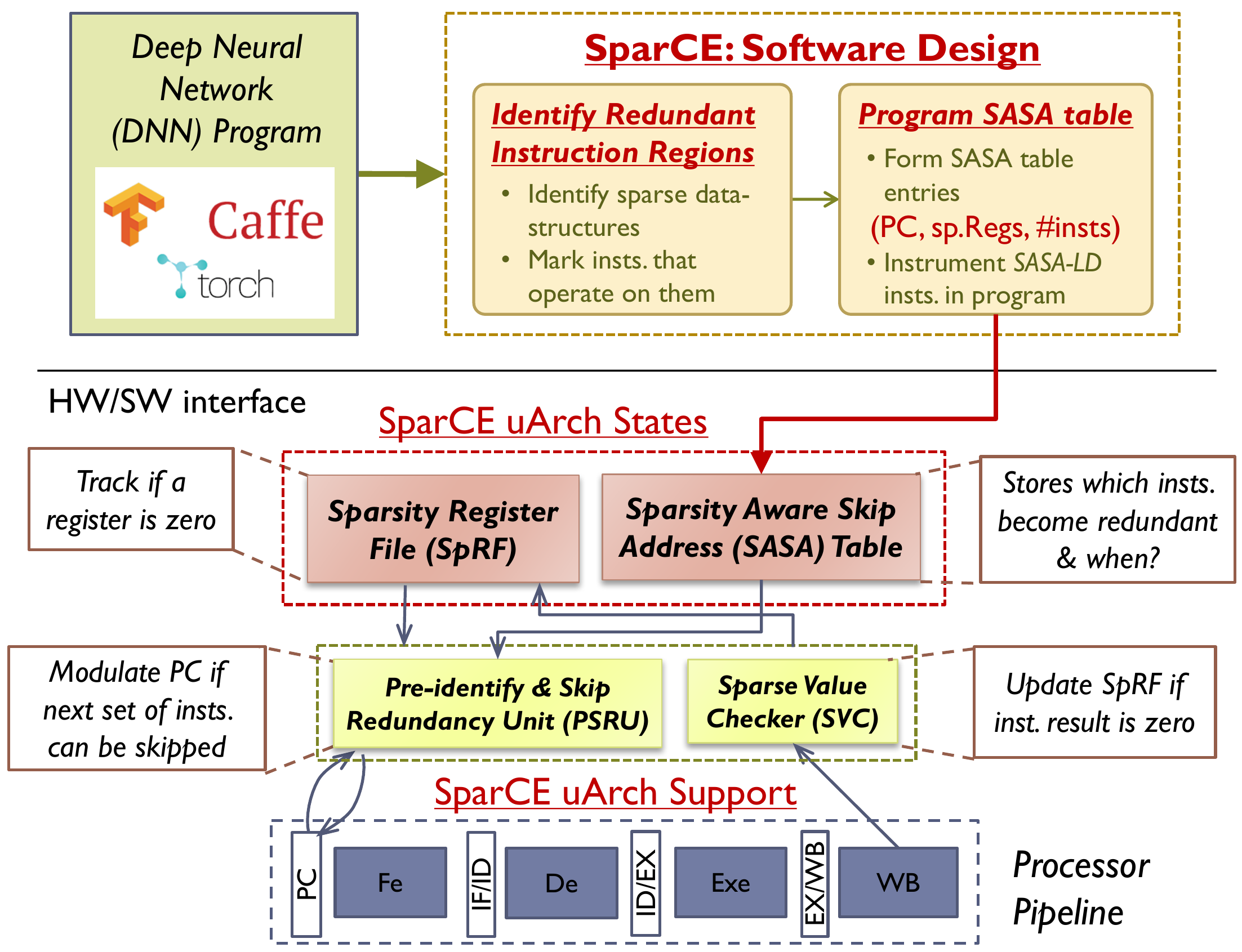}
  \vspace{-20pt}
  \caption{\sagpp: Design Overview}
  \label{fig:sagpp_overview}
\end{figure}

Figure~\ref{fig:sagpp_overview} shows an overview of the micro-architectural and ISA enhancements proposed in \sagpp. We describe these extensions in detail, and demonstrate how they address the aforementioned challenges to leverage sparsity in DNNs. 

\subsubsection{Micro-architectural states and ISA Extension} 
In \sagpp, we augment the processor with two new micro-architectural states \emph{viz.} Sparsity Register File (SpRF) and Sparsity Aware Skip Address (SASA) table. The SpRF is used to dynamically track which registers in the processor's register file contain zero values. The SpRF contains one entry (few bits) corresponding to each register in the register file. When an instruction that writes to a register retires, the SpRF is updated if the result is zero. The SASA table stores information about which instructions can be skipped and under what conditions. Specifically, each entry stores the program counter ($PC$) of the instruction preceding a redundant instruction sequence, the index of the register that determines redundancy and the length of the sequence. Storing the $PC$ of the instruction preceding a redundant instruction sequence allows \sagpp~to \emph{pre-identify whether the next set of instructions can be skipped and if so, skip them before the instructions are fetched}. 

\vspace{3pt}
{\bf \noindent SASA-LD Instruction.} \sagpp~enables software to explicitly identify potentially redundant instruction regions by pre-loading the SASA table at program startup, or before the program execution enters a given code region. To this end, we extend the ISA with a new instruction \emph{viz.} $SASA$-$LD$, which loads a given region of memory into the SASA table. As shown in Equation~\ref{eq:newISA},  the $SASA$-$LD$ instruction takes a register operand ($Rn$) that points to the SASA table's location in memory and an immediate operand ($size$) that denotes the size of the SASA table. 
\begin{equation}
  \centering
  \label{eq:newISA}
  SASA\text{-}LD \quad [Rn], \#size
\end{equation}
At a given point in the program execution, the number of entries in the SASA table limits the number of redundant instruction regions that can be skipped by \sagpp. However, the SASA table can be periodically refreshed from memory as the program execution progresses. In the context of DNNs, we found that 20 entries in the SASA table suffice to capture all redundant instruction sequences, since the computational kernels are captured by a small number of library ({\em e.g.}, BLAS) functions.

\subsubsection{Tracking, Pre-identifying, and Skipping Redundant Instructions} 
\sagpp~utilizes the SpRF and the SASA table to dynamically skip redundant instructions borne out of sparsity in the input data-structures. As shown in Figure~\ref{fig:sagpp_overview}, the micro-architecture of \sagpp~is extended to support the following functions.

\vspace{3pt}
{\bf \noindent Track Sparse Registers.} \sagpp~contains a Sparse Value Checker (\emph{SVC}), which in the processor's writeback stage compares the result of each instruction to zero and if so, updates the entry corresponding to the instruction's destination register in the SpRF. 

\vspace{3pt}
{\bf \noindent Pre-identify \& Skip Redundant Instructions.}  \sagpp~is equipped with a Pre-identify and Skip Redundancy Unit (PSRU) that utilizes the SASA table to identify and skip redundant instruction regions. For each instruction, we check if its $PC$ contains an entry in the SASA table. An entry in the SASA table indicates that the instruction following the current instruction is the start of a potentially redundant instruction sequence. In this case, the PSRU checks the SpRF to identify if the registers indicated in the SASA table entry are currently zero. If so, it increments the $PC$ to the end of the redundant instruction sequence, thereby skipping instructions to benefit performance. If not, \sagpp~proceeds to execute instructions in program order.

In summary, \sagpp~uses the SpRF and the SASA table to seamlessly track sparse registers, pre-identify instruction sequences that are redundant and dynamically skip them before they are even fetched to improve performance.

\begin{figure*}[htb]
  \centering
  \includegraphics[clip,width=\textwidth]{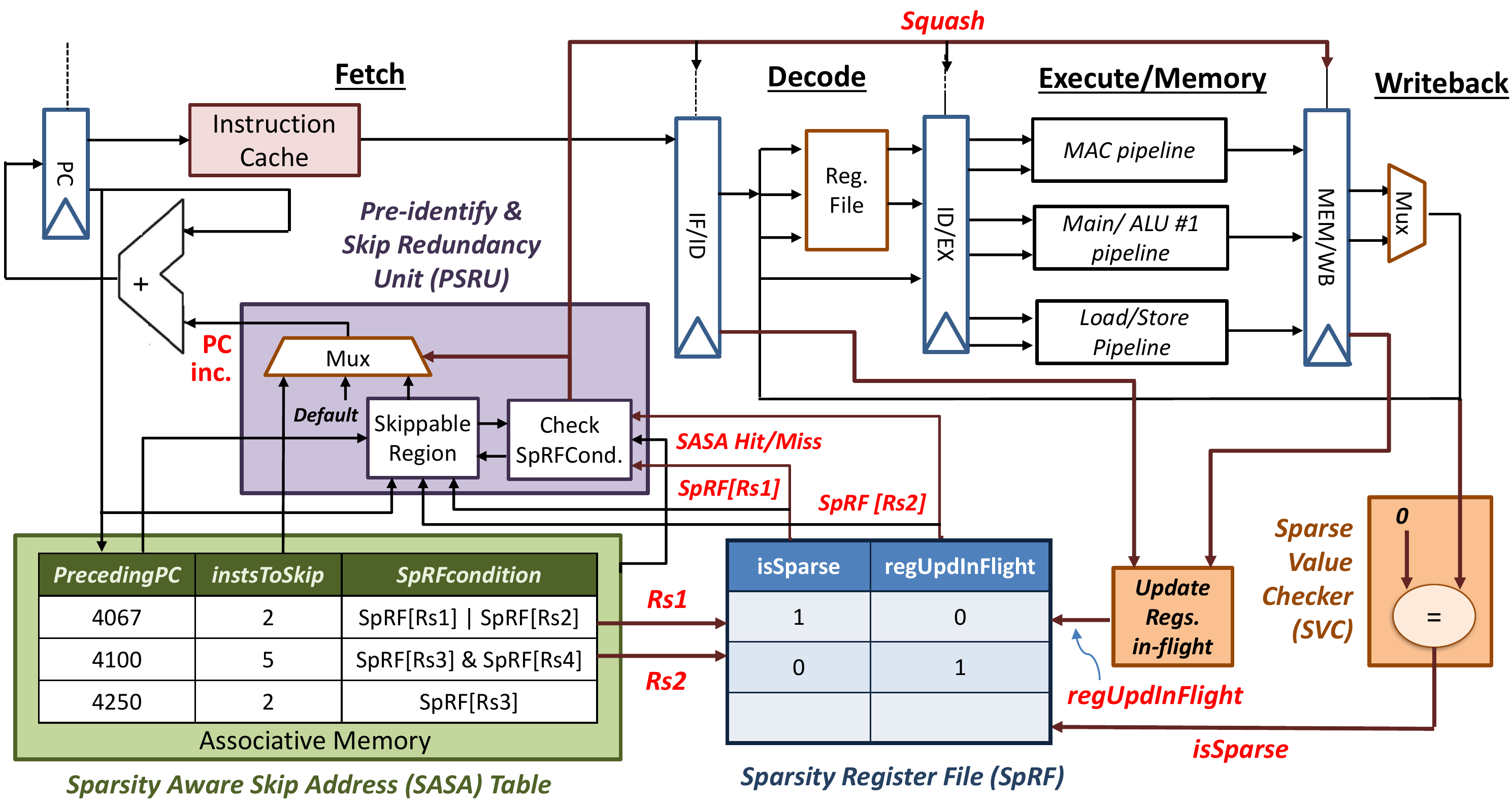}
  \vspace*{-20pt}
  \caption{Block diagram of \sagpp~in-order processor architecture}
  \label{fig:pipeline}
  \vspace*{-15pt}
\end{figure*}

\subsection{In-order SparCE Processor Pipeline} \label{subsec:sapp}
We now explain how \sagpp~can be integrated into an in-order processor. Figure~\ref{fig:pipeline} shows the block diagram of the overall \sagpp~processor architecture. We start with a conventional 4-stage (fetch, decode, execute/memory, and writeback) pipelined processor architecture implementing a RISC-style instruction set with at most 2 source register operands and one destination register operand. Although the \sagpp~architecture is described in this section with a scalar execution unit for ease of illustration, it is directly applicable to vector processors with any number of SIMD execution lanes. We augment the processor with the following structures. 

\vspace{5pt}
\noindent \textbf{Sparsity Register File (SpRF).} The SpRF is located at the fetch stage of the \sagpp~processor. It is a multi-ported register file, with one entry corresponding to each register in the processor's register file. Each entry in the SpRF contains only two single-bit fields - \emph{isSparse} and \emph{regUpdInFlight}. The \emph{isSparse} bit is set to 1 for registers containing zeros, and reset otherwise. The \emph{regUpdInFlight} bit indicates whether an instruction modifying the register is in flight in any stage of the pipeline. For instance, an instruction modifying register $Rd$, sets the $SpRF[Rd][regUpdInFlight]$ field when it enters the decode stage and resets it after committing its result in the writeback stage. In determining whether a future instruction is redundant, the \emph{regUpdInFlight} field ensures that we do not use a stale \emph{isSparse} value when a more recent instruction updating this register is under execution. 

\vspace{5pt}
\noindent \textbf{Sparse Value Checker (SVC).} The SVC is added to the writeback stage of the \sagpp~processor. It contains a comparator that checks if the output of each instruction that updates a register is zero. It then correspondingly updates the SpRF. For example, when the output of an instruction with destination register $Rd$ is zero, then $SpRF[Rd][isSparse]$ is set and $SpRF[Rd][regUpdInFlight]$ is reset. 

\vspace{5pt}
\noindent \textbf{Sparsity Aware Skip Address (SASA) table.} The SASA table is also present in the fetch stage of the \sagpp~processor. As shown in Figure~\ref{fig:pipeline}, the SASA table is an associative memory structure with three fields: (i) $preceedingPC$, which stores the $PC$ of the instruction prior to the redundant instruction sequence, (ii) $SpRFCondition$ field which stores a Boolean combination of 2 register indices in the SpRF that should be satisfied for the instruction region to be skipped, and (iii) $instsToSkip$, which contains the length of the redundant instruction sequence. As an example, for the code in Figure~\ref{fig:loop}, the SASA table entry to skip instructions 6-7 would be \{$preceedingPC$=5, $SpRFCondition$=$r0|r1$, $instsToSkip$=2\}

\begin{figure}[htb]
  \centering
  \includegraphics[clip,width=0.8\columnwidth]{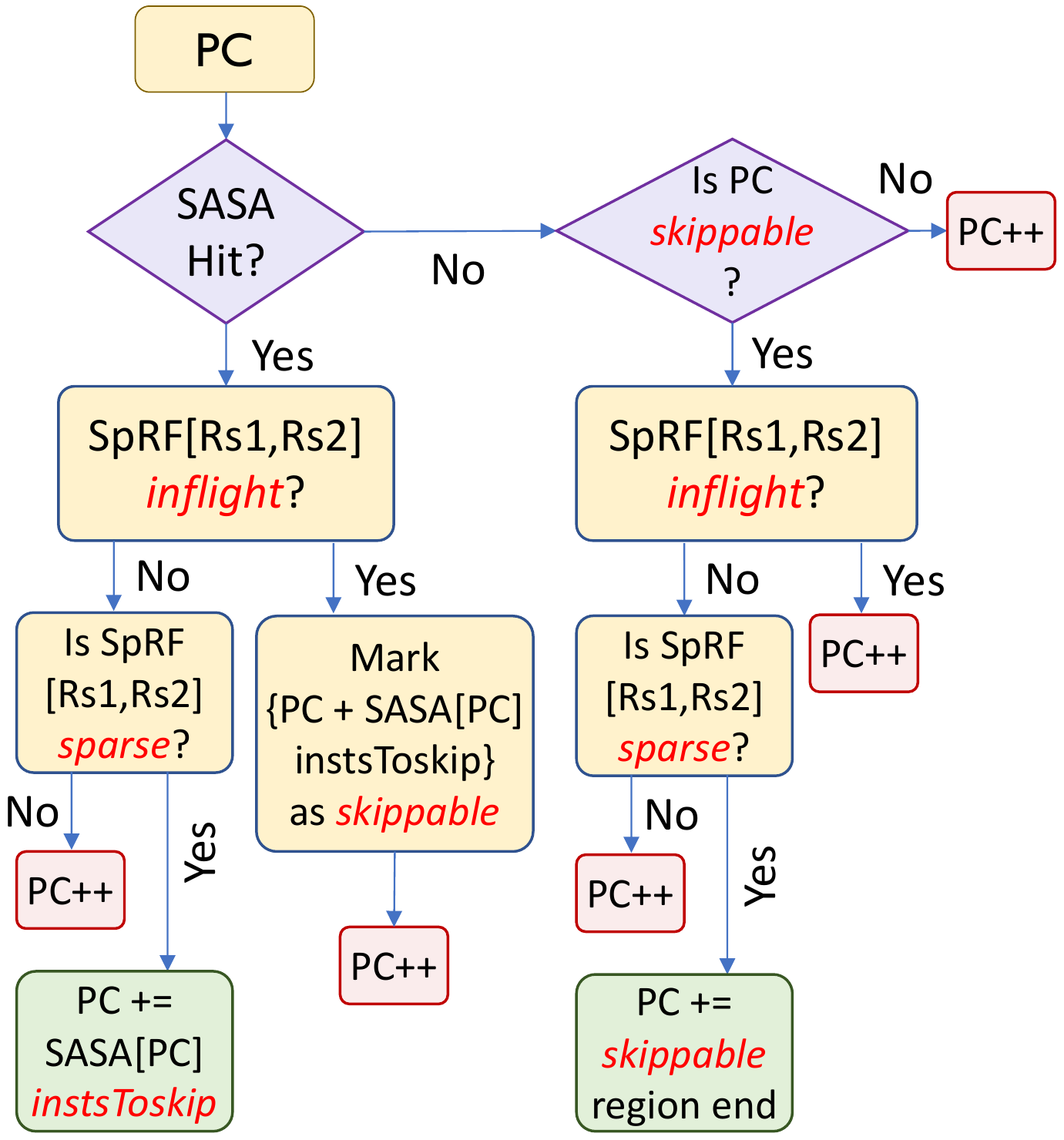}
  \caption{Flowchart for pre-identifying and skipping redundancy}
  \label{fig:flowchart}
\end{figure}

\vspace{5pt}
\noindent \textbf{Pre-identify and Skip Redundancy Unit (PSRU)} The PSRU is also added to the fetch stage of \sagpp~processor. The PSRU determines whether an instruction region specified in the SASA table can be skipped, and appropriately modulates the $PC$. Figure~\ref{fig:flowchart} shows the flowchart depicting the operation of PSRU. Given a $PC$, an associative lookup is performed on the $preceedingPC$ field of the SASA table. If the $PC$ is a hit in the SASA table, then the next instruction marks the beginning of a potentially redundant instruction sequence. To ascertain if the instruction sequence can be skipped, the PSRU reads the register indices specified in the $SpRFCondition$ field of the SASA table (say $R_{s1}$ and $R_{s2}$) from the SpRF. If neither $SpRF[R_{s1}]$ nor $SpRF[R_{s2}]$ have their $regUpdInFlight$ field set, then PSRU computes the Boolean condition (specified in the $SpRFCondition$ field) on their respective $isSparse$ fields. If the condition is satisfied, then the instruction region is deemed redundant and the $PC$ is incremented by $instsToSkip$ to point to the instruction immediately following the end of the redundant region. If not, $PC$ is incremented by 1 and the instructions are executed in program order.
However, if the $regUpdInFlight$ field is set for either $SpRF[R_{s1}]$ or $SpRF[R_{s2}]$ and the Boolean condition in $SpRFCondition$ cannot be definitively evaluated, then the instruction region is temporarily marked as a \emph{skippable} region within the PSRU. The $PC$ is incremented by 1 and the program execution is continued. It is worth noting that continuing or aborting execution of a skippable region will \emph{not} affect program functionality.

In the case when a $PC$ is not present in the SASA table, the PSRU checks if the $PC$ is part of an active skippable region. For the registers present in the $SpRFCondition$, the $regUpdInFlight$ fields are re-checked from the SpRF. If they are reset, the Boolean condition in $SpRFCondition$ is evaluated. If the skippable region is determined to be redundant, then the remaining instructions in the region are skipped by appropriately modifying the $PC$. Further, instructions belonging to the skippable region prior to the current $PC$ are squashed if they are still in flight. If the Boolean condition evaluates to a \emph{false}, then the remaining instructions in the skippable region are executed. Finally, for a $PC$ that misses the SASA table and is not part of any active skippable region, the $PC$ is incremented by 1 to fetch the next instruction.

We note that the logic introduced in \sagpp~to pre-identify redundant instructions executes in parallel with the instruction cache (ICache) access. The additional logic does not impact the latency of the fetch stage, as both the SASA table and the SpRF are significantly smaller structures compared to the ICache.

In summary, by using the SpRF and the SASA table, \sagpp~dynamically tracks sparse registers, pre-identifies if an instruction region is redundant and skips instructions before they are even fetched to improve performance. Thus \sagpp~enables DNN acceleration on GPPs by exploiting the sparsity resident in their data-structures.

\section{Software for SparCE \\ Processsors}
\label{sec:codeGen}
\begin{figure*}[htb]
  \centering
  \vspace*{-10pt}
  \includegraphics[clip,width=\textwidth]{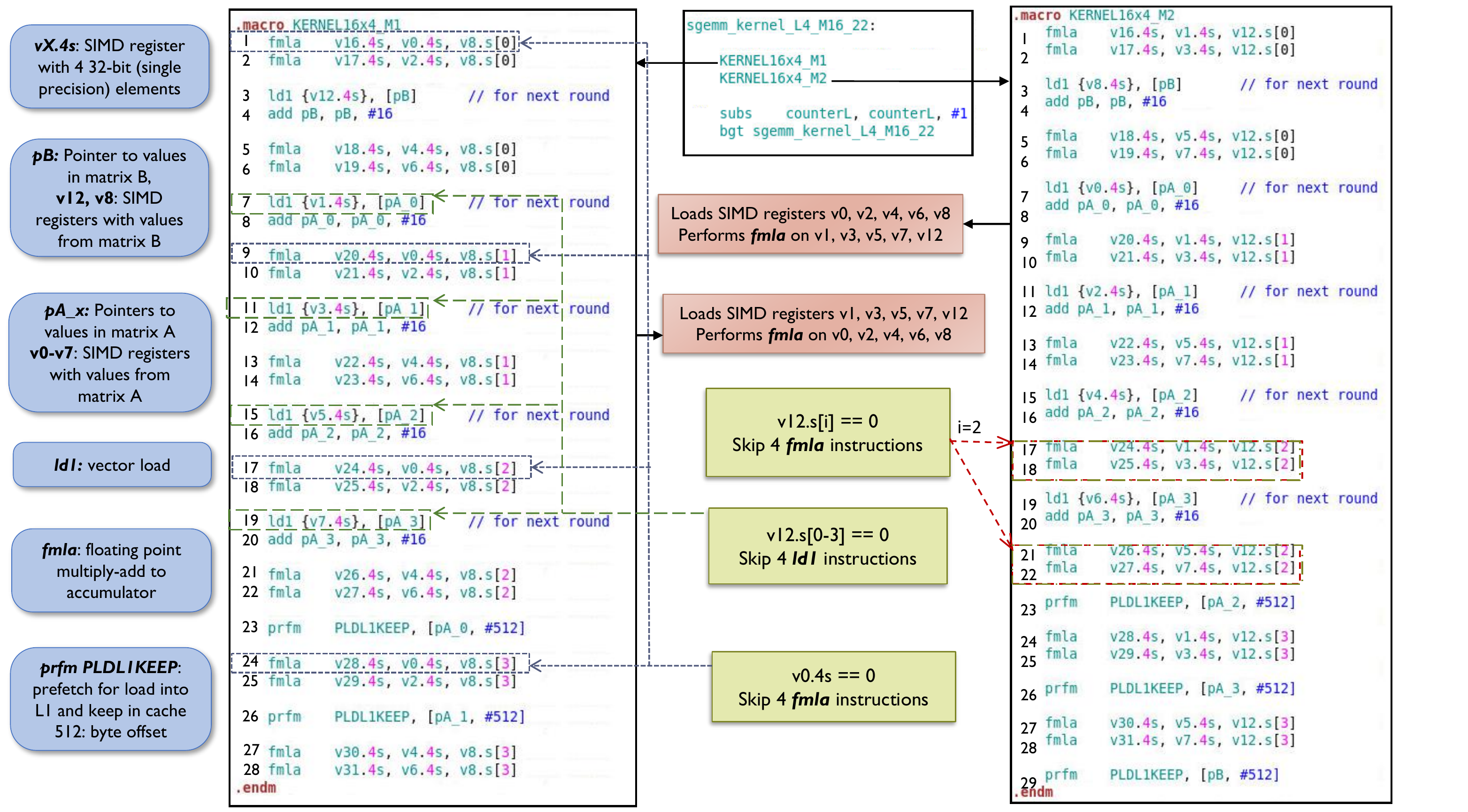}
  \vspace*{-20pt}
  \caption{Zero skipping for \emph{sgemm\_kernel} subroutine in BLAS}
  \label{fig:blas_code}
  \vspace*{-15pt}
\end{figure*}

To extract maximum performance from \sagpp, the software needs to suitably leverage the sparsity-aware micro-architectural extensions. In this section, we outline the key principles behind software design for \sagpp, and demonstrate them in the context of a highly optimized implementation of matrix multiplication (GEMM) from the OpenBLAS library.

\subsection{SparCE Software Design Steps}
The following steps need to be performed in software to leverage sparsity on \sagpp~processors.

\vspace{3pt}
\noindent \textbf{Identifying sparse data-structures and redundant instructions.} One of the key requirements on software is to indicate (through the SASA table) which instruction regions are potentially redundant due to sparsity. To this end, sparse data-structures in the workload need to be first identified by the programmer. Next, when the application is compiled, the registers that hold the sparse data-structures and the instructions that load them from memory are identified. Then, a static dependency analysis of the instruction stream reveals the instructions that are affected by the sparse data-structures, which are then marked as potentially redundant instruction sequences. Following this, the condition for skipping each instruction sequence is derived from: which sparse data-structures affect the region, the registers containing them and the type of operation performed. Finally, the application assembly is instrumented with appropriate $SASA$-$LD$ instructions.

\vspace{3pt}
\noindent \textbf{Separate redundant instructions from instructions causing redundancy.} To identify if an instruction region can be skipped, the instruction whose result makes them redundant should have completed execution. Therefore, in the context of in-order processors, they need to be spaced at least few instructions (3 in our as case as SVC is located in writeback stage) apart in the program, so that redundant instructions are not fetched and no pipeline bubble is introduced due to squashing redundant instructions. To this end, the instruction stream is re-ordered by inserting independent non-redundant instructions where needed to enable sufficient separation.

\vspace{3pt}
\noindent \textbf{Mapping sparse data-structures on vector processors.} In vector processors, typically one of the input operands is shared by all SIMD lanes, while the other is different across lanes. A vector instruction can be skipped only if computations performed on all SIMD lanes are redundant. Hence, it is better to map a sparse data-structure as the shared input operand, since the likelihood of all non-shared inputs being zero is low. If both data-structures are sparse, then the data-structure that exhibits the most block-wise sparsity is mapped as the non-shared input operand.

\subsection{Case Study: SparCE GEMM Routine} \label{sec:sparce_gemm}
Popular deep learning frameworks, such as Caffe~\cite{caffe}, tensor flow~\cite{tensorflow}, \emph{etc.,} execute DNNs as a series of matrix multiply operations, and leverage highly optimized software libraries such as BLAS (Basic Linear Algebra Subprograms) to realize them. Therefore, we develop a \sagpp~version of the GEMM (Generalized Matrix Multiply) routine from the OpenBLAS library~\cite{openblas} targeting an ARM processor. We utilize the \sagpp-GEMM routine to quantify the reduction in DNN execution time achieved on a \sagpp~processor.

Figure~\ref{fig:blas_code} shows the assembly program for the \emph{sgemm\_kernel\_l1\_M16\_22} subroutine, which performs single-precision floating point matrix multiplication ($B$$\times$$A$$=$$C$). The \emph{sgemm} routine executes two smaller subroutines \emph{viz.} \emph{kernel16x4\_M1} (which we abbreviate as $M1$) and \emph{kernel16x4\_M2} ($M2$) sequentially in a loop. The subroutines utilize vector registers $v8$ and $v12$ to hold the first input operand ($B$) and registers $v0$-$v7$ to hold the second operand ($A$). The intermediate results are computed in registers $v16$-$v31$. To optimize performance, each subroutine prefetches data for the other. For example, $M1$ fetches data for operand $A$ into the \emph{odd} registers, which are then used by $M2$ in the subsequent iteration. The memory addresses are provided by scalar registers $pB$ and $pA\_0$-$pA\_3$. The subroutines also prefetch data in the L1-cache using the \emph{prfm} instruction.

We identify redundant instruction sequences in the \emph{sgemm\_kernel\_l1\_M16\_22} subroutine assuming one of the input operands (say $B$) is sparse. The analysis can be easily extended to cover the scenario when either $A$ or both $A$ and $B$ are sparse. Since $B$ is sparse, it is beneficial to map it as the operand shared across the SIMD lanes, which is already the case in Figure~\ref{fig:blas_code}. Even when one of the words in a vector register containing $B$ ($v8$ or $v12$) is zero, 4 $fmla$ instructions can be skipped. For instance, when $v12.s[1]$ equals $0$, instructions 9,10,13,14 in $M2$ can be skipped. This forms 2 redundant instruction sequences (9-10 and 13-14), each of size 2. This amounts to 16 $fmla$s being skipped when the entire vector register ($v8$ or $v12$) is zero. It is worth noting that, had the sparse data-structure ($B$) been mapped as the non-shared SIMD operand in the program, only 4 $fmla$s could be skipped even when the entire vector register is zero.

In addition to the $fmla$ instructions being skipped, when the vector register is fully zero, the load instructions for the second operand can also be skipped. For example, when $v12$ is zero, $ld1$ instructions (7, 11, 15 and 19) for operand $A$ can be skipped in $M1$. Also, note that we did not re-order any instruction in the \emph{sgemm} routine, as the redundant instruction sequences were sufficiently spaced apart from the instruction that triggers their skipping. In the context of $fmla$ instructions, the vector loads happen in a different subroutine owing to pre-fetching, and in the case of the $ld1$ instructions, they were naturally spaced $>$3 instructions apart.

\begin{figure}[htb]
  \centering
  \includegraphics[clip,width=0.8\columnwidth]{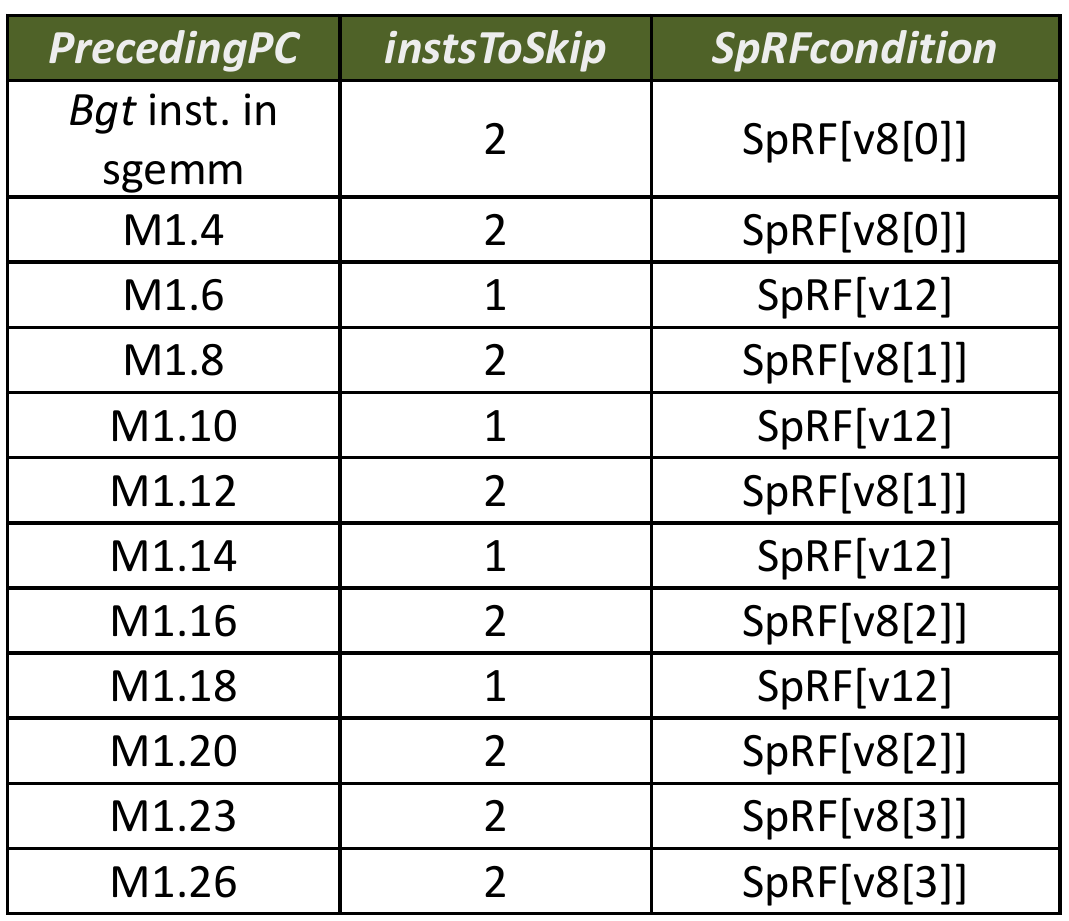}
  \caption{SASA table entries for \emph{kernel16x4\_M1} subroutine}
  \label{fig:sasa_m1}
  \vspace*{-15pt}
\end{figure}

Based on the above analysis, Figure~\ref{fig:sasa_m1} shows the SASA table corresponding to the $M1$ subroutine. We find a total of 12 entries in the SASA table, 8 entries of size 2 for the 16 $fmla$s and 4 entries of size 1 for the $ld1$ instructions.

\begin{figure}[htb]
  \centering
  \includegraphics[clip,width=\columnwidth]{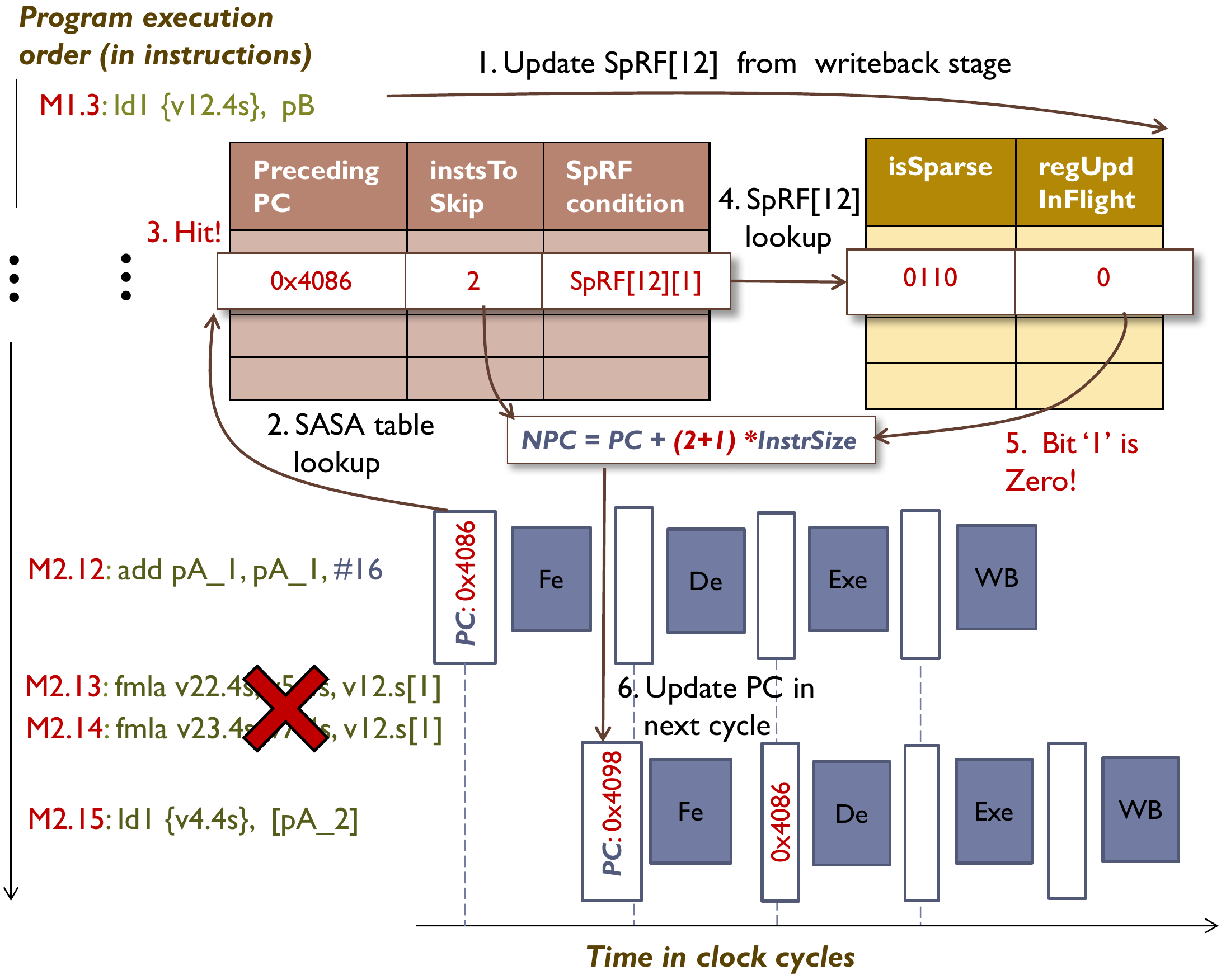}
  \caption{\sagpp~in action for \emph{sgemm} routine}
  \label{fig:blas_action}
  \vspace*{-10pt}
\end{figure}

\vspace{3pt}
{\bf \noindent SparCE in action.} Figure~\ref{fig:blas_action} depicts the sequence of events that leads to instructions being skipped by \sagpp. First, when register $v12$ is loaded by the $M1$ subroutine, its SpRF entry is updated. Note that the $isSparse$ field is a bit vector, one bit for each word in the vector register. Next, when instruction $M2.12$ is fetched, it sees a hit in the SASA table and the reads $SpRF[12]$ to ascertain if the bit corresponding to $SpRF[12][1]$ is sparse. Since that is the case, the $PC$ is incremented to directly fetch $M2.15$.

In summary, with minimal changes to software, \sagpp~processors can leverage sparsity to improve performance.

\vspace*{-0pt}
\section{Experimental Methodology}
\label{sec:exptMeth}
\noindent
In this section, we present the methodology adopted in our experiments to evaluate \sagpp.

\subsection{Performance Evaluation}
We modeled the micro-architectural extensions proposed in \sagpp~using the cycle-accurate \emph{gem5} architectural simulator~\cite{gem5}. The gem5 simulator was tightly integrated with the popular Caffe~\cite{caffe} deep learning framework, wherein the matrices corresponding to each layer and each input batch was formed in Caffe and fed into the gem5 simulator to perform the matrix computations. The results were fed back to Caffe to form the inputs for the next layer (or input batch), and so on. Figure~\ref{fig:table}(a) shows the gem5 system configuration used in our experiments. All experiments were run in full-system mode. We evaluate \sagpp~by measuring application level execution times under two scenarios. The first scenario targets embedded scalar processors that are present in ultra-low power edge/IoT devices and lack support for high performance libraries. We chose a scalar ARM v8 in-order processor architecture as the baseline. We leverage the modular nature of gem5 to cater to this scenario, wherein we disable support for advanced architectural features such as SIMD processing and pre-fetching in the ARM v8 processor. We then prototyped a direct convolution routine (which we call \emph{Dir-Conv-Scalar}) that does not utilize the disabled features and used it in our experiments. The second scenario targets a reasonably sophisticated mobile processor for which we chose the ARM v8 in-order processor architecture with 4-way SIMD as the baseline. In this case, the matrix computations were realized using the highly-optimized OpenBLAS~\cite{openblas} based GEMM routines described in Section~\ref{sec:sparce_gemm}. We refer to this implementation as \emph{OpenBLAS-SIMD4}.

\subsection{Power and Area Evaluation}
\begin{figure}[htb]
  \centering
  \includegraphics[clip,width=1.0\columnwidth]{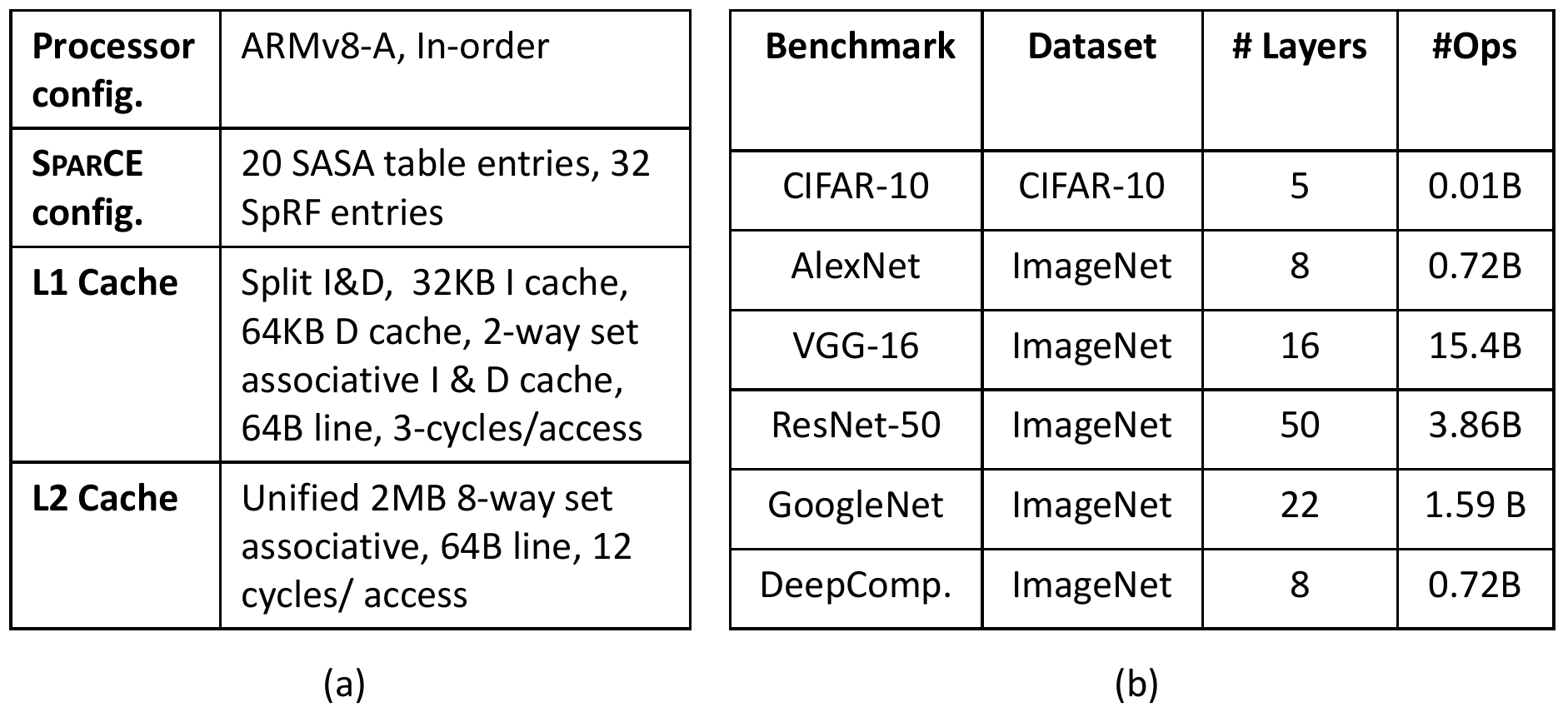}
  \caption{(a) Gem5 simulation parameters (b) Application benchmarks}
  \label{fig:table}
\end{figure}

We implemented the hardware extensions of \sagpp~at the Register Transfer Level (RTL) using Verilog HDL and synthesized them to IBM 45nm technology using Synopsys Design Compiler to measure its power and area overheads. For the configuration shown in Figure~\ref{fig:table}(a), the area overhead is 0.4\% of the ARM Cortex A35 core~\cite{arm}. Thus the area overhead of \sagpp~is quite minimal allowing its deployment in the resource-constrained embedded platforms.

\subsection{Benchmarks}
Our benchmark suite, listed in Figure~\ref{fig:table}(b), consists of 6 state-of-the-art image-recognition DNNs \emph{viz.} CIFAR-Caffe DNN using the CIFAR-10 dataset~\cite{cifarData}, and AlexNet~\cite{alexNet}, VGG-16~\cite{vgg16}, ResNet-50~\cite{resNet}, GoogleNet~\cite{googleNet} and Deep Compression-AlexNet~\cite{deepComp} using the ImageNet dataset. These benchmarks contained 5-50 layers and took 0.01-15.4 Billion scalar operations to classify an image. We utilized pre-trained models from the Caffe Model Zoo to evaluate \sagpp~in the context of inference. For training, we utilized only the smaller CIFAR-10 benchmark, as training ImageNet models on gem5 was prohibitively time consuming. It is noteworthy that all benchmarks exhibited dynamic sparsity in features and errors, while only Deep Compression-AlexNet exhibited static parsity in weights.

\section{Results}
\label{sec:results}
\noindent
In this section, we present the results of our experiments that highlight the advantages of \sagpp.

\subsection{Performance and Energy Improvement}
\begin{figure}[htb]
  \centering
  \includegraphics[clip,width=1.0\columnwidth]{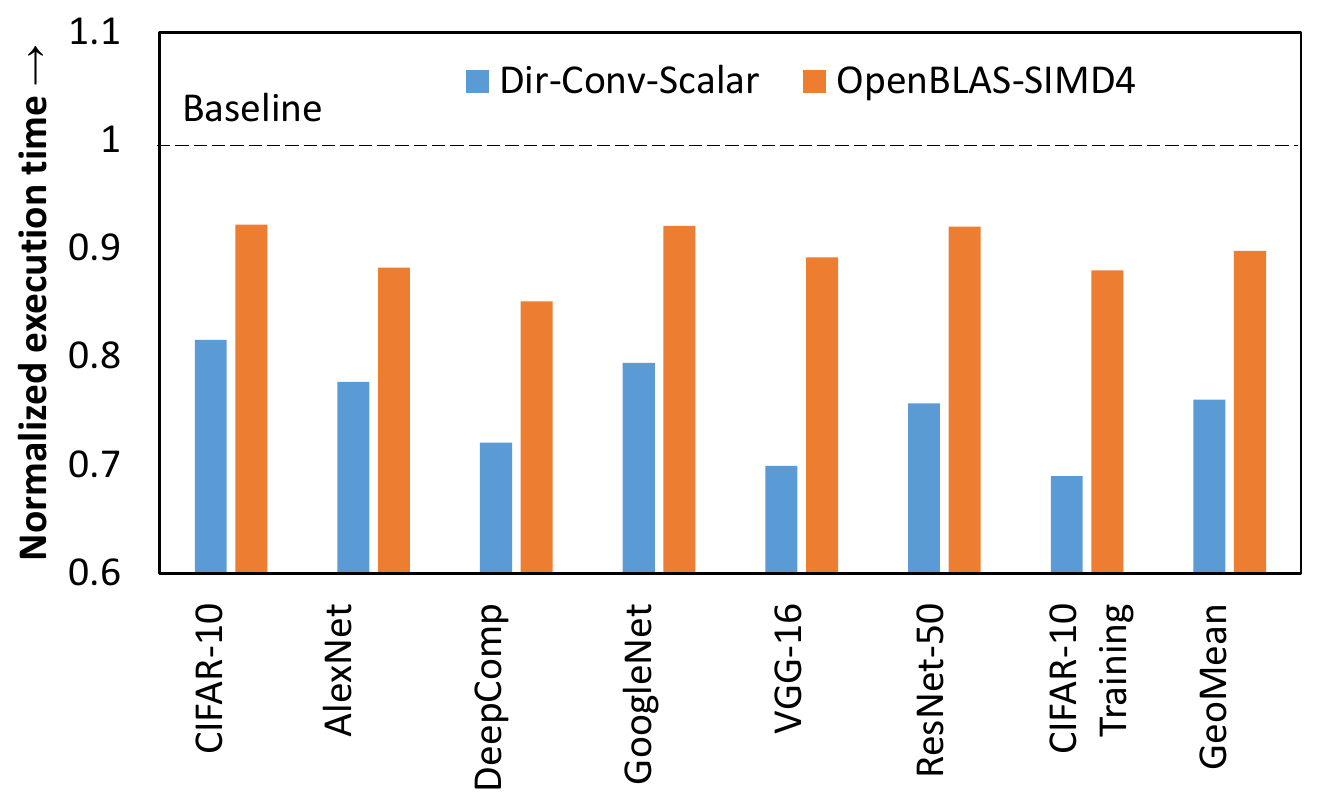}
  \caption{Improvement in execution time at the application level}
  \label{fig:timeBenefits}
\end{figure}

Figure~\ref{fig:timeBenefits} shows the normalized execution time benefits of \sagpp~over the baseline processor for both inference and training. In the context of Dir-Conv-Scalar, the reduction in \emph{application runtime} ranges between 19\%-31\% across the benchmarks. In contrast, OpenBLAS-SIMD4 demonstrates benefits in the range of 8\%-15\% reduction in runtime. This is because $fmla$ instructions occupy a much smaller fraction of their runtime, as their execution engines are more sophisticated - multiple SIMD lanes, low floating point instruction latency \emph{etc.} Also, since they  support features such as prefetching where the data is already fetched into the higher levels of the memory subsystem, avoiding redundant data fetches has a less prominent impact on performance.

Among the benchmarks, the execution time benefits are largely proportional to the amount of sparsity that they exhibit (Figure~\ref{fig:src_sparsity} in Section~\ref{sec:motivation}). The Deep Compression-AlexNet benchmark achieves the most benefits because both its feature and weight data-structures are sparse, as opposed to other benchmarks whose weight data-structure is dense. In the context of training, the backpropagation step achieves more improvement compared to forward propagation. This stems from the fact that the error data-structure is more sparse compared to features.

\begin{figure}[htb]
  \centering
  \includegraphics[clip,width=\columnwidth]{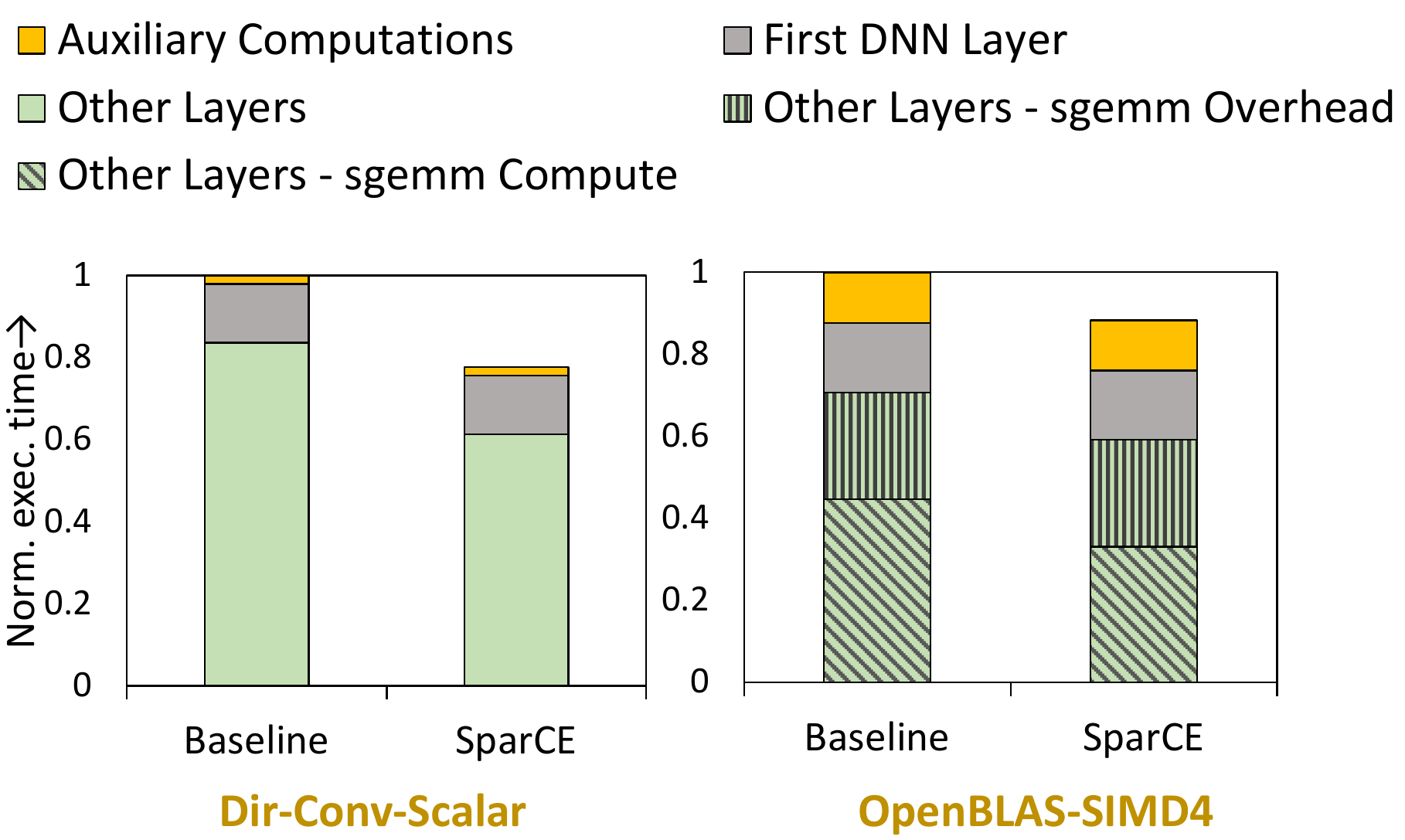}
  \vspace{-10pt}
  \caption{Execution time breakdown for AlexNet}
  \label{fig:time_brk}
  \vspace{-5pt}
\end{figure}

\vspace{3pt}
{\bf \noindent Execution Time Breakdown.} To better appreciate the improvements achieved by \sagpp, Figure~\ref{fig:time_brk} hierarchically breaks down the application execution time (in the context of both Dir-Conv-Scalar and OpenBlas-SIMD4 implementations) into components that can and cannot be impacted by \sagpp. At the top level, the solid yellow and gray colors represent the execution time fraction that cannot be improved by \sagpp. This is primarily constituted by auxiliary DNN operations such as activation functions, subsampling and others, which represent 1.9\% and 12.2\% of the runtime for Dir-Conv-Scalar and OpenBlas-SIMD4 implementations, respectively. It is noteworthy that although these operations occupy $<$1\% of the total DNN FLOPs, they occupy a substantially larger fraction of the runtime for the OpenBLAS-SIMD4 implementation. This is owed to the fact that they are typically memory-bound (higher Bytes/FLOP ratio), which is further amplified as matrix multiply operations are significantly optimized by the GEMM subroutine. Also, since the inputs to the DNN are typically dense, the first DNN layer exhibits little redundancy. This occupies 14.3\% and 16.9\% of the total runtimes of AlexNet for Dir-Conv-Scalar and OpenBLAS-SIMD4 implementations, respectively. The fraction grows smaller in deeper networks such as ResNet and VGG.

In Figure~\ref{fig:time_brk} the green color bars denote the computations that can accelerated by leveraging sparsity ($\sim$71\%). For Dir-Conv-Scalar implementation, this is limited to 83.6\% of the baseline AlexNet runtime. Since AlexNet contains $\sim$36\% redundant computations (Figure~\ref{fig:bm_opportunity} in Section~\ref{sec:motivation}), \emph{the best case benefits are limited to 29.8\%}, of which \sagpp~achieves 22.3\%. For the OpenBLAS-SIMD4 implementation, the underlying GEMM involves supplementary operations like memory allocate, copy and free operations, which as marked by the dotted patterns and consumes $27\%$ of the total execution time. This constraints the opportunity for \sagpp~to $\sim$44\% of AlexNet runtime as shown by the diagonally hatched portions in Figure~\ref{fig:time_brk}. Since AlexNet contains $\sim$36\% redundant computations (Figure~\ref{fig:bm_opportunity} in Section~\ref{sec:motivation}), \emph{the best case benefits are limited to $\sim$16\%}, of which \sagpp~achieves 12\% improvement as other control operations such as pointer arithmetic, prefetching \emph{etc.} present within the loop body cannot be avoided.

\begin{figure}[htb]
  \centering
  \includegraphics[clip,width=0.9\columnwidth]{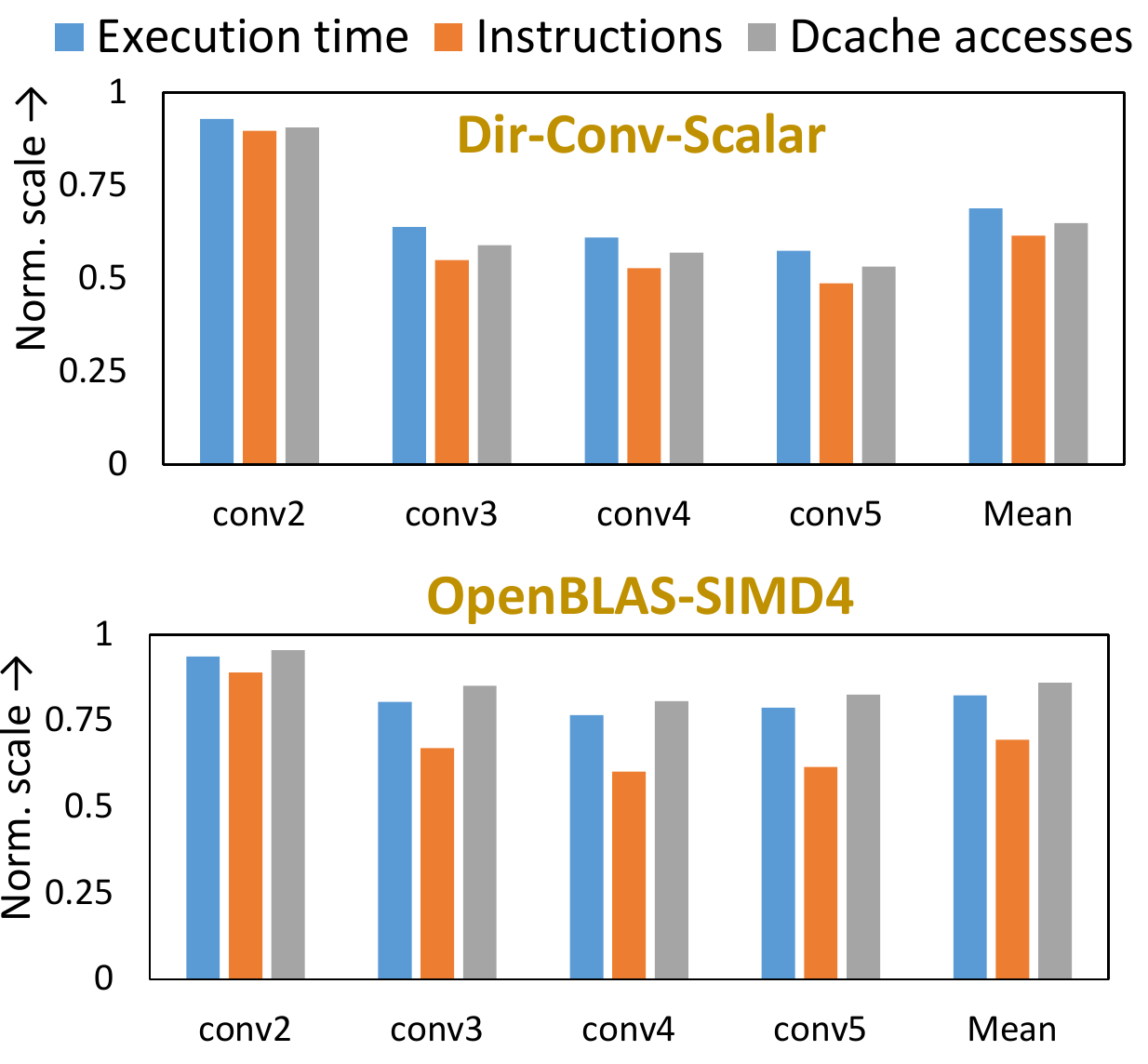}
  \vspace{-10pt}
  \caption{Layer-wise benefits breakdown for AlexNet}
  \label{fig:brk1}
\end{figure}

{\bf \noindent Layer-wise Benefits.} We now present the layer-wise breakdown of the benefits quantified in terms of the execution time, instructions and data cache (D-Cache) accesses skipped for the convolutional layers of AlexNet. Figure~\ref{fig:brk1} shows the benefits in the context of both SIMD and scalar processor implementations. As shown in Figure~\ref{fig:brk1}, we are able to achieve, on an average, 39.4\% reduction in instruction count and 35.1\% reduction in D-Cache accesses for Dir-Conv implementations on low-power embedded scalar processors. The reduction in instruction count and D-Cache accesses amount to 30.5\% and 14\% respectively for OpenBLAS-SIMD4 implementations. We also observe the benefits are typically larger for layers deeper in the DNN, as they typically exhibit more sparsity.

\vspace{3pt}
{\bf \noindent Energy Benefits.} Power evaluation of the \sagpp~reveals that it consumes 1.74 mW at 1 GHz, which amounts to 1.9\% of the 90 mW power consumed by even the most power-efficient baseline ARM v8 processor, the Cortex A35 processor~\cite{arm}. Accordingly, the execution time benefits translate to benefits in the range of 16.9\%-28.7\% reduction in application-level energy for a Dir-Conv-Scalar implementation. In the context of OpenBLAS-SIMD4 implementations, the reduction in energy ranges between 6.1\%-13.2\% across the benchmarks.

\subsection{Performance Scaling with Sparsity}
\begin{figure}[htb]
  \centering
  \includegraphics[clip,width=1.0\columnwidth]{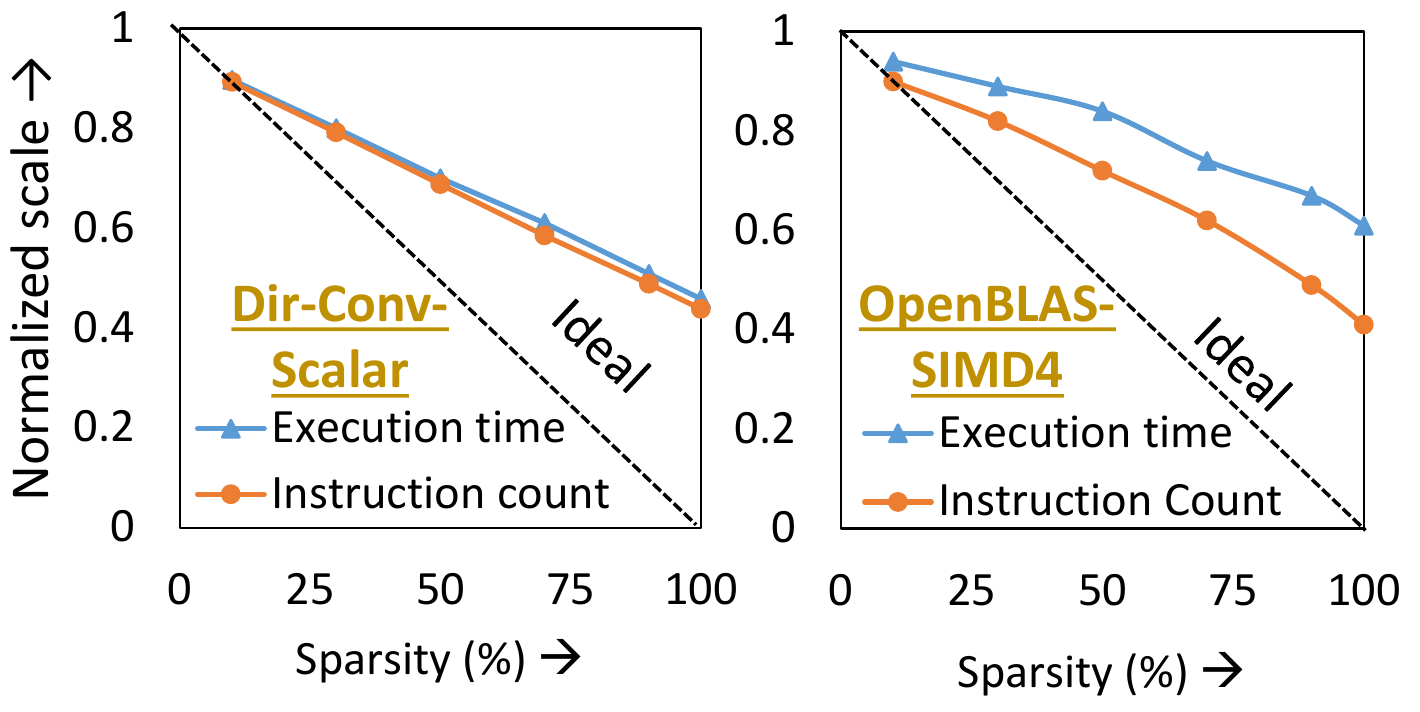}
  \vspace*{-10pt}
  \caption{\sagpp~performance scaling with sparsity}
  \label{fig:varSparsity}
\end{figure}

We now study how the performance of \sagpp~scales with increasing levels of sparsity. To this end, we considered a matrix multiplication problem ($B$$\times$$A$$=$$C$), wherein the dimensions of the input matrices $B$ and $A$ were 169$\times$3456 and 3456$\times$384 respectively. 
We varied the sparsity of the $B$ matrix by constraining the number of zero entries. The location of the zeros and other entries of the matrices were chosen at random. Figure~\ref{fig:varSparsity} shows how the execution time and the fraction of instructions executed varies with sparsity in the context of both Dir-Conv-Scalar and OpenBLAS-SIMD4 implementations. We find that both implementations exhibit strong performance scaling with sparsity, outlining the ability of \sagpp~to efficiently skip computations. We find the number of instructions executed to be larger than ideal (dotted line in Figure~\ref{fig:varSparsity}) due to the presence of control instructions for pointer arithmetic, loop counts \emph{etc.,} in the program, which cannot be skipped. Also, the disparity in the fraction of instructions executed and the resultant execution time benefits is more pronounced for the OpenBLAS-SIMD4 implementation. We attribute this to the intelligent instruction ordering in the GEMM routine utilized in the implementation, wherein computations are aggressively overlapped with data-fetches. Therefore, even if computations are skipped, the improvement in performance is limited by the time taken for the data-fetches.

\subsection{Operand Ordering in SparCE OpenBLAS-SIMD4 Implementations}
\begin{wrapfigure}{r}{0.5\columnwidth}
  \centering
  \includegraphics[clip,width=0.5\columnwidth]{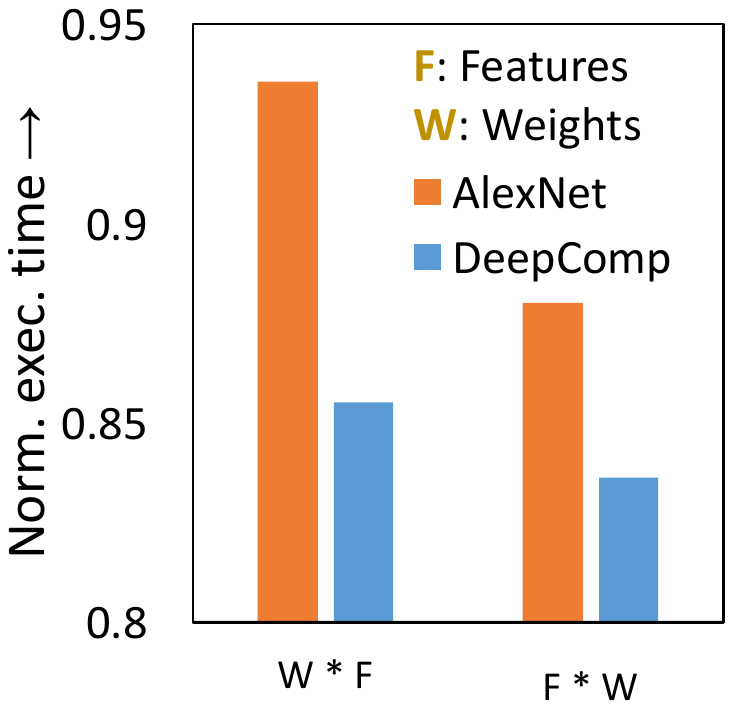}
  \caption{Impact of operand ordering on performance}
  \label{fig:deepComp}
\end{wrapfigure}

As described in Section~\ref{sec:codeGen}, based on the amount of sparsity exhibited by the data-structures, mapping the right data-structure as the shared-SIMD operand can have a considerable impact on performance. In the case of all benchmarks other than Deep Compression-AlexNet, only the feature data-structure is sparse. Therefore, mapping it as the shared-SIMD operand (Matrix $B$ in the \emph{sgemm} subroutine) would yield the best benefits. Figure~\ref{fig:deepComp} shows the performance improvement achieved when operands are ordered in both ways \emph{viz.} Features$\times$Weights and Weights$\times$Features. In the context of AlexNet, we find that mapping features as the shared-SIMD operand yields 1.86$\times$ better benefits ($\sim$12\% \emph{vs.} 6.5\%) compared to mapping the non-sparse weight data-structure.
 For the Deep Compression-AlexNet network, since both feature and weight data-structures are sparse, the disparity in performance due to operand ordering is relatively small ($<$2\%). Even in this case, we find that choosing features as the shared-SIMD operand is beneficial. This is attributed to the fact that some of the weight matrices have high degree of sparsity, and their zero entries are typically clustered. Therefore, using weights as the non-shared SIMD operand has less of an adverse impact on performance compared to using features. 

\vspace*{-0pt}
\section{Related Work}
\label{sec:relatedwork}
\noindent
Prior research efforts that target improving the computational efficiency of DNNs can be grouped into the following broad classes.

\vspace{3pt}
\noindent \textbf{Software parallelization on multi-cores and GPUs.} A large number of previous efforts have been directed towards developing techniques for efficient parallelization of DNNs on multi-core servers and GPUs~\cite{Krizhevsky:2014, Iandola:2015, Dean:2012, Rhu:2016, Das:2016, Zlateski:2016, Nervana:2016}. However, the scalability of these techniques is often limited by synchronization and communication bottlenecks.

\vspace{3pt}
\noindent \textbf{Specialized accelerators.} Developing specialized hardware architectures has been an attractive approach to improve the computational efficiency of DNNs. These accelerators~\cite{Chen:2014,Chen:2014:1,Venkataramani:2017,Jouppi:2017,Majumdar:2012,Gokhale:2014,Farabet:2011,eyeriss,Chakradhar:2010, Reagen:2016, Eldridge:2015} utilize specialized processing cores, interconnect network and other hardware-software co-design methodologies to leverage the different forms of compute and data reuse patterns in DNNs.

\vspace{3pt}
\noindent \textbf{Approximate computing.} DNNs and the applications that use them are intrinsically resilient to errors in their underlying computations. Approximate computing techniques, such as low-precision implementations~\cite{Courbariaux:2015,Gupta:2015,Zhu:2016,Venkataramani:2014}, model compression~\cite{deepComp,Jaderberg:2014} and others~\cite{Seide:2014,Zhang:2015:1}, leverage this property to improve the computational efficiency of DNNs. 

\vspace{3pt}
\noindent \textbf{Post-CMOS technology.} Post-CMOS technologies such as memristors and spintronics have succeeded in realizing the computational primitives of DNNs through their intrinsic device characteristics. Implementations of small scale DNNs using memristor crossbar arrays~\cite{Liu:2015,Shafiee:2016} and spintronic devices~\cite{Ramasubramanian:2014} have demonstrated significant promise.

All the above efforts are complementary to \sagpp, as unlike above efforts, we attempt to leverage sparsity in the different DNN data-structures to improve efficiency on cost constrained embedded platforms comprising of mainly a GPP core.

\vspace{3pt}
\noindent \textbf{Exploiting sparsity.} Prior efforts that exploit sparsity to improve DNN efficiency can be grouped to two classes based on whether they exploit static sparsity or dynamic sparsity, as shown in Figure~\ref{fig:relwork}. Specialized sparse architectures that are capable of exploiting both static and dynamic sparsity incorporate a variety of compression techniques and zero-skipping schemes to reduce storage requirement and avoid redundant multiplications in accelerators~\cite{cnvlutin, eyeriss, eie, scnn, cambriconx, Park:2016}. 

On the other hand, software approaches that exploit weight sparsity on GPPs take advantage of sparse matrix libraries. These sparse libraries usually yield performance improvement only under extreme levels of sparsity ($>$95\%). Since DNNs naturally exhibit sparsity in the range of 40\%-70\%, a few research efforts force more weight sparsity into DNNs using sparse decomposition methods, \emph{etc.}~\cite{BLiu:2015,Wen:2016} or customize the pruning to match the underlying hardware organization~\cite{Yu:2017}. These invariably come at the cost of training overhead or loss of functional accuracy. In contrast, we propose micro-architectural extensions to GPPs that can exploit both dynamic and static sparsity while being effective even under intermediate levels. Thus, \sagpp~is able to exploit feature sparsity in several state-of-the-art dense DNN models as well as both feature and weight sparsity in existing pruned DNN models.    

\section{Conclusion}
\label{sec:conclusion}
\noindent
As DNNs pervade the spectrum of computing devices, new approaches to improve their computational efficiency on resource-constrained IoT/ edge devices becomes critical. In this work, we accelerate DNNs on GPPs, which are an indispensable part of IoT/ edge devices, by exploiting sparsity in the different DNN data-structures. To this end, we propose sparsity aware general purpose core extensions (\sagpp) that enable GPPs to efficiently leverage sparsity, while being minimally intrusive and low-overhead. \sagpp~comprises of two key micro-architectural enhancements. First, a Sparsity Register File (SpRF) dynamically tracks zero-valued registers in the processor. Next, a Sparsity Aware Skip Address (SASA) table indicates potentially redundant instruction sequences and the conditions under which they can be skipped. A Pre-identify and Skip Redundancy Unit (PSRU) combines the information from the SpRF and the SASA table to dynamically \emph{pre-identify} if an instruction sequence can be skipped, and if so masks it from being fetched and executed. We evaluate \sagpp~on 6 image-recognition DNNs in the context of both training and inference. Our evaluations reveal that \sagpp~is a promising design that allows us to exploit all forms of static and dynamic sparsity to accelerate DNNs on GPPs. 


\newpage
\bibliographystyle{unsrt}
\bibliography{ref}

\end{document}